\documentclass[usenatbib]{mn2e}
\usepackage{graphicx, subfig, tabularx, bm, amssymb, aas_macros, amsmath, natbib}
\usepackage[normalem]{ulem}
\usepackage[colorlinks=true, citecolor=blue, urlcolor=blue, linkcolor=black]{hyperref}
\newcommand{\Tspace}{\rule{0pt}{2.6ex}}
\newcommand{\Bspace}{\rule[-1.2ex]{0pt}{0pt}}

\newcommand{\vcirc}{V_{\rm{circ}}}

\newcommand{\vmax}{V_{\rm{max}}}
\newcommand{\rmax}{r_{\rm{max}}}
\newcommand{\mhalf}{M_{1/2}}
\newcommand{\rhalf}{r_{1/2}}
\newcommand{\mhalo}{M_{\rm halo}}

\newcommand{\msub}{M_{\rm{sub}}}

\newcommand{\mvir}{M_{\rm{vir}}}

\newcommand{\vvir}{V_{\rm{vir}}}
\newcommand{\tvir}{T_{\rm{vir}}}
\newcommand{\mthree}{M_{300}}
\newcommand{\msix}{M_{600}}

\newcommand{\mstar}{M_{\star}}
\newcommand{\msun}{M_{\odot}}

\newcommand{\lsun}{L_{\odot}}

\newcommand{\mpc}{{\rm Mpc}}
\newcommand{\kpc}{{\rm kpc}}
\newcommand{\kms}{{\rm km \, s}^{-1}}

\newcommand{\zacc}{z_{\rm infall}}
\newcommand{\vacc}{V_{\rm infall}}
\newcommand{\macc}{M_{\rm infall}}

\newcommand{\lcdm}{$\Lambda$CDM}
\DeclareBoldMathCommand{\bmlcdm}{\Lambda{\rm CDM}}

\newcommand{\estar}{\epsilon_{\star}}
\newcommand{\bkbk}{BBK}

\DeclareBoldMathCommand{\bvfifty}{V<50\,\kms}

\title[The Milky Way's bright satellites in \lcdm]
{
The Milky Way's bright satellites as an apparent failure of $\bmlcdm$ 
}
\author[M. Boylan-Kolchin, J. S. Bullock and M. Kaplinghat]
{Michael Boylan-Kolchin,\thanks{$\!$Center for Galaxy Evolution
  fellow}\thanks{email: m.bk@uci.edu} James S. Bullock and Manoj Kaplinghat\\
\noindent $\!\!$Center for Cosmology, Department of Physics and Astronomy, 
  4129 Reines Hall, University of California, Irvine, CA 92697, USA}
\begin{document}

 \pagerange{\pageref{firstpage}--\pageref{lastpage}} 
 \pubyear{2012}

\maketitle

\label{firstpage}
\begin{abstract}
  We use the Aquarius simulations to show that the most massive subhalos in
  galaxy-mass dark matter halos in \lcdm\ are grossly inconsistent with the
  dynamics of the brightest Milky Way dwarf spheroidal galaxies.  While the
  best-fitting hosts of the dwarf spheroidals all have $12 \la \vmax \la 25 \,
  \kms$, \lcdm\ simulations predict at least ten subhalos with $\vmax >
  25\,\kms$.  These subhalos are also among the most massive at earlier times,
  and significantly exceed the UV suppression mass back to $z \sim 10$.  No
  \lcdm-based model of the satellite population of the Milky Way explains this
  result.  The problem lies in the satellites' densities: it is straightforward
  to match the observed Milky Way luminosity function, but doing so requires the
  dwarf spheroidals to have dark matter halos that are a factor of $\sim 5$ more
  massive than is observed.  Independent of the difficulty in explaining the
  absence of these dense, massive subhalos, there is a basic tension between the
  derived properties of the bright Milky Way dwarf spheroidals and \lcdm\
  expectations.  The inferred infall masses of these galaxies are all
  approximately equal and are much lower than standard \lcdm\ predictions for
  systems with their luminosities.  Consequently, their implied star formation
  efficiencies span over two orders of magnitude, from 0.2\% to 20\% of baryons
  converted into stars, in stark contrast with expectations gleaned from more
  massive galaxies.  We explore possible solutions to these problems within the
  context of \lcdm\ and find them to be unconvincing.  In particular, we use
  controlled simulations to demonstrate that the small stellar masses of the
  bright dwarf spheroidals make supernova feedback an unlikely explanation for
  their low inferred densities.
\end{abstract}


\section{Introduction} 
\label{sec:intro}
It has been over ten years since it was first pointed out that the number of
satellite galaxies of the Milky Way is much smaller than the number of dark
matter subhalos predicted by cold dark matter (CDM)-based simulations of Milky
Way-mass systems \citep{klypin1999, moore1999}.  These subhalos formed as
independent dark matter halos and were subsequently accreted onto the Milky
Way's halo, so each of them is a potential site of galaxy formation; that we
only see a small number of Milky Way satellites therefore requires explanation.
This discrepancy may originate from galaxies populating only a subset of the
predicted subhalos, or from modifications of the CDM model on small scales such
that many of these dark matter subhalos do not actually exist.  Understanding
which (if either) solution is correct will have vastly different implications.

In this paper, we focus on the former possibility, which associates the bright
satellites of the Milky Way with the most massive subhalos expected from \lcdm\
simulations.  These models can be further split into two broad classes: those
that link a satellite's luminosity to the present day mass of its dark matter
halo \citep{stoehr2002, hayashi2003, penarrubia2008}, and those that assume
satellites form only in the biggest halos defined at some earlier epoch, e.g.,
reionization \citep{bullock2000, kravtsov2004, ricotti2005, koposov2009,
  okamoto2009}.  Building on our recent work \citep*{boylan-kolchin2011a} that
has demonstrated a puzzling discrepancy between the masses of massive simulated
subhalos and the dynamics of bright Milky Way satellites, we perform a thorough
analysis of the consistency of models associating bright satellites with massive
dark matter subhalos.  We show that neither of these classes reconciles
observations with \lcdm\ simulations, absent further modifications: massive
subhalos, defined either now or at any earlier epoch, are generically too dense
to be dynamically consistent with the Milky Way satellites.  This problem
relates to the structure of subhalos, making it distinct from the standard
missing satellites problem, which is usually phrased in terms of abundances.

In fact, galaxy abundance issues are not limited to the Milky Way: it has been
known for decades that the dark matter halo mass function has a steep low-mass
slope while the faint end of the galaxy luminosity function is flat \citep{press1974,
  schechter1976}, signifying a broad need for a non-linear mapping between
galaxy luminosity and halo mass in cold dark matter models (e.g.,
\citealt{white1978}).  In this sense, the missing satellites problem can be
interpreted as the low-luminosity manifestation of the faint galaxy issue that
has been the focus of significant effort in galaxy formation theory throughout
the modern era.

In recent years, many groups have shown that a simple statistical association
between dark matter halos and galaxies can be used to reproduce the clustering
of galaxy light and mass, and its evolution with redshift, using only dark
matter structures extracted from $N$-body simulations (e.g.,
\citealt{kravtsov2004a, conroy2006, moster2010, guo2010}).  By enforcing the
cumulative abundance of galaxies to be equal to that of halos -- $n(>\mhalo) =
n(>\mstar)$ -- one obtains a direct galaxy-halo relation that has been tested to
scales as small as $\mhalo \sim 5\times 10^{10}\,\msun$ \citep{blanton2008}.
Abundance matching also apparently works well at reproducing the luminosity
function of satellite galaxies in the Milky Way \citep{busha2010, kravtsov2010,
  lunnan2012}, engendering optimism that the missing satellites problem can be
solved naturally in \lcdm\ by a $\mstar(\mhalo)$ relation that falls steeply
towards low halo masses.  The problem would then be reduced to understanding
what causes a strong suppression of galaxy formation in low-mass halos, and
there are at least two well-motivated processes that likely contribute to this
inefficiency.

The first crucial component is thought to be supernova feedback.  If even a
relatively small amount of the energy from supernovae can couple to the
interstellar medium of a galaxy, it can remove gas from shallow gravitational
potential wells and thereby regulate star formation in low-mass systems
\citep{larson1974, dekel1986}.  The second process is reionization of the
Universe.  After the Universe is reionized, the resultant UV background affects
the gaseous content of dark matter halos both by photo-evaporation of existing
gas and by suppressing subsequent accretion through an increase in the
temperature of the intergalactic medium \citep{efstathiou1992, thoul1996,
  barkana1999, gnedin2000, hoeft2006, okamoto2008}.  Combined, these effects
reduce the baryon fraction of halos after reionization in a time-varying manner,
such that more and more massive halos are affected as time progresses.
Photo-heating and photo-evaporation therefore may have an important effect on
galaxy formation in the host halos of the Milky Way's dwarf galaxies
\citep{bullock2000, benson2002a, somerville2002, kravtsov2004, ricotti2005,
  madau2008, bovill2009, busha2010, lunnan2012}.

Semi-analytic models of galaxy formation have shown that supernova feedback and
reionization should indeed strongly limit the formation of stars in low-mass
halos since $z \approx 6$ \citep{benson2002a, koposov2009, munoz2009,
  li2010, maccio2010, guo2011, font2011}.  While the detailed abundance and chemical
composition of MW satellites produced in such models depends on the precise
combination of reionization and supernova feedback \citep{font2011}, it is
nevertheless encouraging that such models can reproduce many properties of the
MW's satellite population using well-established physics.  Models and
simulations including inherently inefficient star formation (due to the
difficulties in forming molecular gas in low-mass halos) have also reproduced
dwarf galaxy scalings \citep{robertson2008, tassis2008, gnedin2009, kuhlen2011}.
Additional feedback processes such as cosmic ray pressure may also help suppress
star formation in low mass halos \citep{wadepuhl2011}.

In light of this substantial progress in modeling and successes in comparing
observations to simulations, it is tempting to consider the \lcdm\ model a
success at reproducing the main properties of the satellite population of the
Milky Way.  At least one substantial issue remains, however.  As we showed in a
recent paper \citep*[hereafter \bkbk]{boylan-kolchin2011a}, the majority of the
most massive subhalos -- the putative hosts of dwarf spheroidals -- in the
highest resolution $N$-body simulations currently available are dynamically
inconsistent with all of the dSphs: massive Milky Way subhalos predicted by
\lcdm\ are too dense to host the bright MW dSphs.  Considering not only the
mass, but also the structure, of the dSphs' dark matter halos provides an
additional constraint beyond pure abundance matching, and shows that while
abundance matching can reproduce the luminosity function of the Milky Way
satellites, it does so at the expense of assigning the bright satellites to
halos that are too dense to actually host the MW dSphs. This surprising result
presents a new challenge for \lcdm\ models on sub-galactic scales.

In this paper, we present a thorough comparison of the bright ($L_V >
10^{5}\,\lsun$) MW dSphs and simulated MW subhalos, complementing and expanding
our results in \bkbk.  A substantial difference between \bkbk\ and this work is
our treatment of subhalo mass profiles: rather than relying on mass models for
the subhalos inferred from their structural properties, we use the full particle
data from the numerical simulations for our comparison.
Section~\ref{section:methods} describes the observational data and numerical
simulations upon which our analysis is based.  We show direct comparisons of
subhalo circular velocity profiles with the observed structure of Milky Way
dSphs in Section~\ref{sec:lcdm_mw}, demonstrating that there is a significant
disagreement between the two.  In Section~\ref{sec:dwarf_dm_mass}, we use
Bayesian inference to determine the masses of the bright dSphs' dark matter
hosts and show that the results are inconsistent with any model relying on an
association of the brightest satellites with the most massive dwarfs (defined at
any epoch).  Section~\ref{sec:discussion} explores possible reasons for the
mismatch between the densities of subhalos and the dynamics of dwarfs.

\vskip0.4cm

\section{Simulations and Data}
\label{section:methods}
\subsection{Simulations}
\label{subsec:simulations}

Our \lcdm\ predictions are based on dark matter halos from the Aquarius project
\citep{springel2008}, which consists of six Milky Way-mass dark matter halos
(denoted A-F) selected at $z\!=\!0$ from a large cosmological
simulation\footnote{This was presented in \citet{gao2008} as the `hMS'
  simulation, and was resimulated at $\sim 10\times$ higher mass resolution as
  the Millennium-II Simulation \citep{boylan-kolchin2009}} and resimulated at a
variety of mass and force resolutions.  The specific halos selected for
re-simulation were chosen at random after applying a mild isolation cut
(additionally, the selected halos are predicted by a semi-analytic model to host
late-type galaxies) and are generally typical of halos with similar mass
\citep{boylan-kolchin2010}.  Only halo A was simulated at the highest resolution
(level 1), in which the particle mass was $m_p=1.7\times 10^{3}\,\msun$ and the
Plummer equivalent gravitational softening length was $\epsilon=20.5\,{\rm pc}$.
All halos were simulated at level 2 resolution, with $m_p=6.4\times 10^{3}-1.4
\times 10^{4}\,\msun$ and $\epsilon=65.8\,{\rm pc}$, resulting in approximately
120 million particles within each halo's virial radius.  These six constitute
our sample of simulated dark matter halos.

The masses of the Aquarius halos are $(0.95-2.2) \times 10^{12}\,\msun$, a range
that reflects the uncertainty in the true value of the MW's virial mass and
covers almost all recent estimates (see Sec.~\ref{subsec:mw_mass} for a more
detailed discussion of various estimates of the mass of the Milky Way's dark
matter halo).  The exact definition of virial mass $\mvir$ itself varies among
different authors: it is defined to be the mass of a sphere, centered on the
halo in question, containing an average density $\Delta$ times the critical
density of the Universe, but different authors adopt different values of
$\Delta$.  Throughout this paper, we use $\Delta=\Delta_{\rm vir}$, the value
derived from the spherical top-hat collapse model \citep{gunn1972, bryan1998},
which results in $\Delta_{\rm vir} \approx 94$ at $z=0$ for the cosmology used
by the Aquarius simulations (see below).  $\Delta=200$ and
$\Delta=200\,\Omega_m(z)$ are two other common values used in the literature.

For each halo, self-bound substructures were identified using the
\texttt{SUBFIND} algorithm \citep{springel2001} as described in more detail in
\citet{springel2008}.  Substructures can be characterized by their total bound
mass $\msub$, or by a characteristic circular velocity $\vmax$, defined to be
the maximum of the circular velocity $\vcirc=\sqrt{GM(<r)/r}$.  We will
typically use $\vmax$ (and $\rmax$, defined as $\vcirc(\rmax)=\vmax$) rather
than $\msub$ when discussing subhalos because $\vmax$ is less dependent on the
specific algorithm used to identify subhalos and compute their properties.

%%%%%%%%%%%%%%%%%%%%%%%%%%%%%%%%%%%%%%%%%%%%%%%%%%%%%%%%%%%% 
\begin{table}
  \caption{
    Properties of the Aquarius simulations.  \textit{Columns}: (1) Simulation;
    (2) virial mass; and
    (3)--(6) number of subhalos at $z=0$ within $300$ kpc of the halo's center
    and having  $\vacc > (20, 30, 40, 50)\,\kms$, respectively.
   \label{table:simulations}
  }
  \begin{tabular*}{\columnwidth}{@{\extracolsep{\fill}} ccrrrr}
    \hline
    \hline
    Name & $M_{\rm vir}$ [$\msun$] & $N_{20}$  & $N_{30}$  & $N_{40}$  & $N_{50}$ \\
    \hline
    Aq-A\Tspace & $2.19 \times 10^{12}$& 105& 33& 15& 6\\
    Aq-B & $9.54 \times 10^{11}$ & 60& 16& 7& 1\\
    Aq-C & $1.99 \times 10^{12}$ & 81& 28& 12& 4\\
    Aq-D & $2.19 \times 10^{12}$ & 111& 31& 15& 10\\
    Aq-E & $1.39 \times 10^{12}$ & 85& 25& 11& 3\\
    Aq-F\Bspace & $1.32 \times 10^{12}$ & 99& 29& 12& 5\\
    % VL-II & $1.73 \times 10^{12}$& 84& 27& 14& 5\\
    \hline
  \end{tabular*}
\end{table}
%%%%%%%%%%%%%%%%%%%%%%%%%%%%%%%%%%%%%%%%%%%%%%%%%%%%%%%%%%%% 

Subhalo catalogs were constructed at each time-step for which the full particle
information was saved, typically 128 snapshots per halo.  The subhalos were
linked across snapshots by merger trees, allowing us to explore the full
evolutionary history for each subhalo in addition to its $z=0$ properties.
Motivated by abundance matching models, we also compute the epoch $\zacc$,
defined to be the redshift at which a subhalo's mass is maximized (typically
just prior to infall onto a larger halo), as well as the circular velocity
$\vacc \equiv \vmax(\zacc)$ and mass $\macc \equiv \mvir(\zacc)$ at that time.
In each simulation, we limit our subhalo sample to those within 300 kpc of the
center of the host halo and having $\vmax(z=0) > 10\,\kms$.
Table~\ref{table:simulations} summarizes some important properties of the
Aquarius halos and their massive subhalos.

The Aquarius simulations were performed in the context of a spatially flat WMAP1
cosmological model, with a matter density of $\Omega_m = \!0.25$, a baryon
density of $\Omega_b=0.045$, reduced Hubble parameter $h = 0.73$, linear power
spectrum normalization $\sigma_8=0.9$, and a spectral index of the primordial
power spectrum of $n_s=1$; this is the same cosmology used in the Millennium and
Millennium-II simulations \citep{springel2005a, boylan-kolchin2009}.  Analysis
of the WMAP7 data indicates that $\sigma_8 =0.81 \pm 0.03$ and $n_s=0.967\pm
0.014$ (\citealt{komatsu2011}, based on their ``WMAP seven-year mean'' values),
both of which are somewhat lower than the values used in the Aquarius
simulations.  While reductions in these parameters may change the properties of
halos predicted by \lcdm, \bkbk\ showed that the Via Lactea II (VL-II)
simulation \citep{diemand2008}, which was run with $\sigma_8=0.74$ and
$n_s=0.951$, predicts very similar structural properties of massive MW
satellites to those in Aquarius.  Note that the $\sigma_8$ and $n_s$ values of
VL-II are actually lower than the WMAP7 values, which strengthens the impact of
this comparison.  A preliminary analysis of one MW-mass halo run using WMAP7
parameters shows that the structure of dark matter subhalos in the updated
cosmology is very similar to that in the Aquarius cosmology (Garrison-Kimmel et
al., in preparation).  Current evidence therefore points to our results being
independent of $\sim 10\%$ changes in cosmological parameters, though it is
certainly desirable to have a large sample of high resolution halos simulated in
the WMAP7 cosmology in order to make the most precise predictions possible.

\subsection{Observational data}
\label{subsec:observations}
Our primary data for each dwarf are the measured de-projected half-light radius
($\rhalf$) and the dynamical mass within this radius ($\mhalf$).
\citet{strigari2007a}, \citet{walker2009} and \citet{wolf2010} have shown that
the dynamical masses of the MW dSphs within a radius comparable to their stellar
extent are well-constrained by kinematic data.  In particular, while the radial
mass profile is sensitive to factors such as the velocity anisotropy, these
uncertainties are minimized at $\rhalf$ \citep{wolf2010}, resulting in a
straightforward and accurate (to approximately 20\%) estimate of $\mhalf$, or
equivalently, $\vcirc(\rhalf)$:
\begin{align}
 \mhalf &=3\,G^{-1} \langle \sigma_{\rm los}^2 \rangle\,\rhalf\,,\label{eq:mhalf}\\
  \vcirc(\rhalf) &= \sqrt{3 \,\langle \sigma_{\rm los}^2 \rangle}\,. \label{eq:vhalf}
\end{align}
The brackets in the above equations refer to luminosity-weighted averages.
While Eqn.~\eqref{eq:mhalf} was derived using the spherical Jeans equation,
\citet{thomas2011} have shown that this mass estimator accurately reflects the
mass as derived from axisymmetric orbit superposition models as well.  This
result suggests that Eqns.~\eqref{eq:mhalf} and \eqref{eq:vhalf} are also
applicable in the absence of spherical symmetry, a conclusion that is also
supported by an analysis of Via Lactea II subhalos \citep{rashkov2012}.

We focus on the bright MW dSphs -- those with $L_V > 10^{5}\,\lsun$ -- for
several reasons.  Primary among them is that these systems have the highest
quality kinematic data and the largest samples of spectroscopically confirmed
member stars to resolve the dynamics at $\rhalf$.  The census of these bright
dwarfs is also likely complete to the virial radius of the Milky Way ($\sim
300\,\kpc$), with the possible exception of yet-undiscovered systems in the
plane of the Galactic disk; the same can not be said for fainter systems
\citep{koposov2008, tollerud2008}.
Finally, these systems all have half-light radii that can be accurately resolved
with the highest resolution $N$-body simulations presently available.

The Milky Way contains 10 known dwarf spheroidals satisfying our luminosity cut
of $L_V > 10^{5}\,\lsun$: the 9 classical (pre-SDSS) dSphs plus Canes Venatici
I, which has a $V$-band luminosity comparable to Draco (though it is
significantly more spatially extended).  As in \bkbk, we remove the Sagittarius
dwarf from our sample, as it is in the process of interacting (strongly) with
the Galactic disk and is likely not an equilibrium system in the same sense as
the other dSphs.  Our final sample therefore contains 9 dwarf spheroidals:
Fornax, Leo I, Sculptor, Leo II, Sextans, Carina, Ursa Minor, Canes Venatici I,
and Draco.  All of these galaxies are known to be dark matter dominated at
$\rhalf$ \citep{mateo1998}: \citet{wolf2010} find that their dynamical
mass-to-light ratios at $\rhalf$ range from $\sim 10-300$.

The Large and Small Magellanic Clouds are dwarf irregular galaxies that are more
than an order of magnitude brighter than the dwarf spheroidals.  The internal
dynamics of these galaxies indicate that they are also much more massive than
the dwarf spheroidals: $\vcirc({\rm SMC}) = 50-60\,\kms$
\citep{stanimirovic2004, harris2006} and $\vcirc({\rm LMC}) = 87 \pm 5\,\kms$
\citep{olsen2011}.  Abundance matching indicates that galaxies with luminosities
equal to those of the Magellanic Clouds should have $\vacc \approx 80-100\,\kms$
(\bkbk); this is strongly supported by the analysis of \citet{tollerud2011}.  A
conservative estimate of subhalos that could host Magellanic Cloud-like galaxies
is therefore $\vacc > 60\,\kms$ and $\vmax > 40\,\kms$.  As in \bkbk, subhalos
obeying these two criteria will be considered Magellanic Cloud analogs for the
rest of this work.

\section{Comparing $\bmlcdm$ subhalos to Milky Way satellites}
\label{sec:lcdm_mw}

\subsection{A preliminary comparison}
\label{subsec:preliminary_comparison}
%%%%%%%%%%%%%%%%%%%%%%%%%%%%%%%%%%%%%%%%%%%%%%%%%%%%%%%%%%%%%%
\begin{figure}
 \centering
 \includegraphics[scale=0.58]{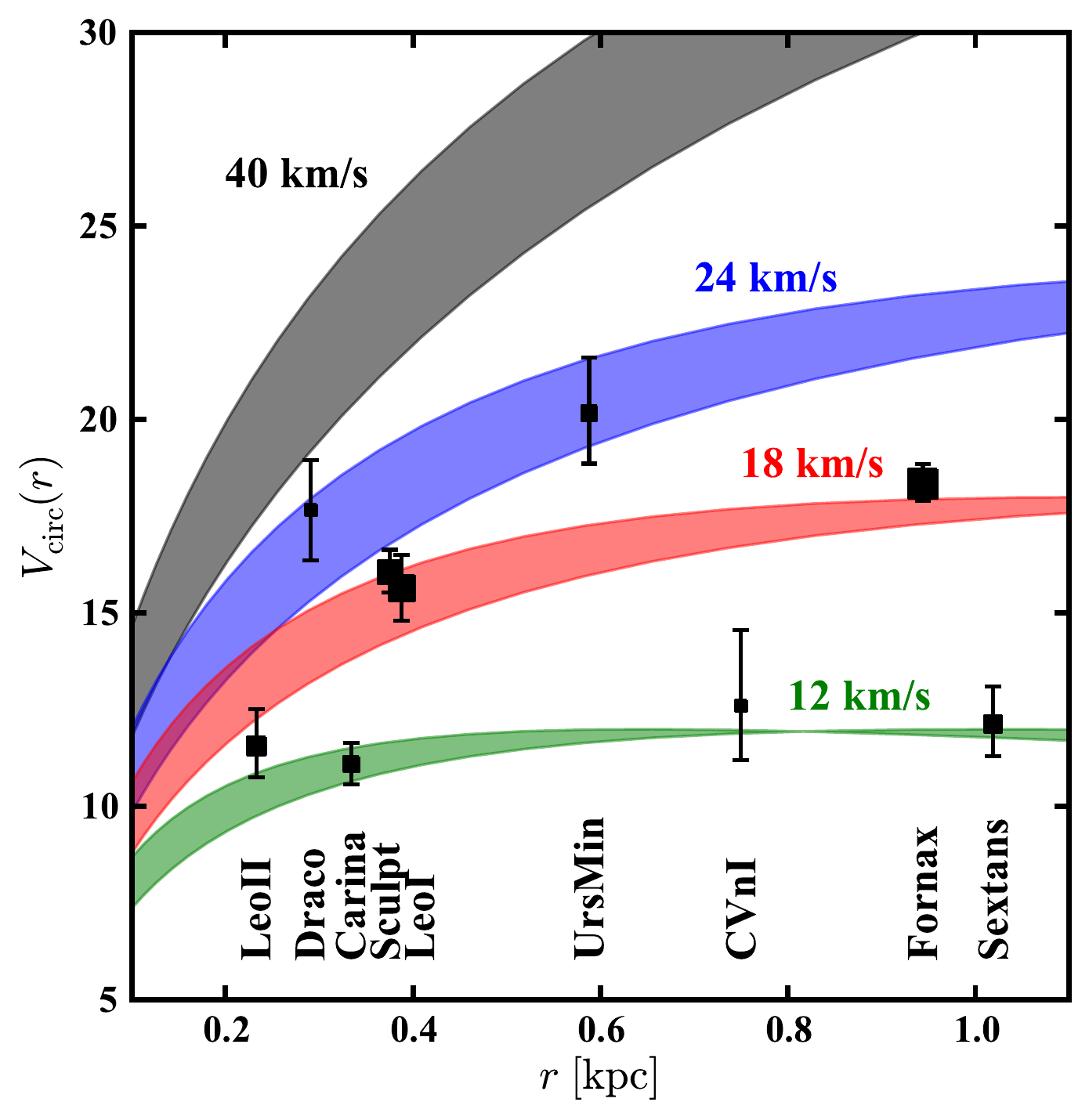}
 \caption{Observed $\vcirc$ values of the nine bright dSphs (symbols, with sizes
   proportional to $\log \, L_V$), along with rotation curves corresponding to
   NFW subhalos with $\vmax=(12, \,18, \, 24, \, 40)\, \kms$.  The
   shading indicates the $1\, \sigma$ scatter in $\rmax$ at fixed $\vmax$ taken
   from the Aquarius simulations.  All
   of the bright dSphs are consistent with subhalos having $\vmax \leq 24
   \,\kms$, and most require $\vmax \la 18\,\kms$.  Only Draco, the least
   luminous dSph in our sample, is consistent (within $2 \sigma$) with a massive
   CDM subhalo of $\approx 40\,\kms$ at $z=0$.
\label{fig:vcirc_theory}
}
\end{figure}
%%%%%%%%%%%%%%%%%%%%%%%%%%%%%%%%%%%%%%%%%%%%%%%%%%%%%%%%%%%%%% 

Density and circular velocity profiles of isolated dark matter halos are
well-described (on average) by \citet[hereafter, NFW]{navarro1997} profiles,
which are specified by two parameters -- i.e., virial mass and concentration, or
$\vmax$ and $\rmax$.  Average dark matter subhalos are also well-fitted by NFW
profiles inside of their tidal radii, though recent work has shown that the
3-parameter Einasto (\citeyear{einasto1965}) profile provides a somewhat better
match to the profiles of both simulated halos \citep{navarro2004, merritt2006b,
  gao2008, ludlow2011} and subhalos \citep{springel2008} even when fixing the
Einasto shape parameter (thereby comparing models with two free parameters
each).  To connect this work to the analysis of BBK,
Figure~\ref{fig:vcirc_theory} compares the measured values of $\vcirc(\rhalf)$
for the nine bright MW dSphs to a set of dark matter subhalo rotation curves
based on NFW fits to the Aquarius subhalos; the shaded bands show the
$1\,\sigma$ scatter from the simulations in $\rmax$ at fixed $\vmax$.  More
detailed modeling of subhalos' density profiles will be presented in subsequent
sections.

It is immediately apparent that all of the bright dSphs are consistent with NFW
subhalos of $\vmax = 12-24 \,\kms$, and only one dwarf (Draco) is consistent
with $\vmax > 24 \, \kms$.  Note that the size of the data points is
proportional to galaxy luminosity, and no obvious trend exists between $L$ and
$\vcirc(\rhalf)$ or $\vmax$ (see also \citealt{strigari2008}).  Two of
the three least luminous dwarfs, Draco and Ursa Minor, are consistent with the
most massive hosts, while the three most luminous dwarfs (Fornax, Leo I, and
Sculptor) are consistent with hosts of intermediate mass ($\vmax \approx
18-20\,\kms$).  Each of the Aquarius simulations contains between 10 and 24
subhalos with $\vmax > 25\,\kms$, almost all of which are insufficiently massive
to qualify as Magellanic Cloud analogs, indicating that models populating the
most massive redshift zero subhalos with the brightest MW dwarfs will fail.

\subsection{Assessing the consistency of massive $\bmlcdm$
  subhalos with bright Milky Way satellites}
\label{subsec:compare_profiles}
%%%%%%%%%%%%%%%%%%%%%%%%%%%%%%%%%%%%%%%%%%%%%%%%%%%%%%%%%%%%%%
\begin{figure*}
 \centering
 \includegraphics[scale=0.46]{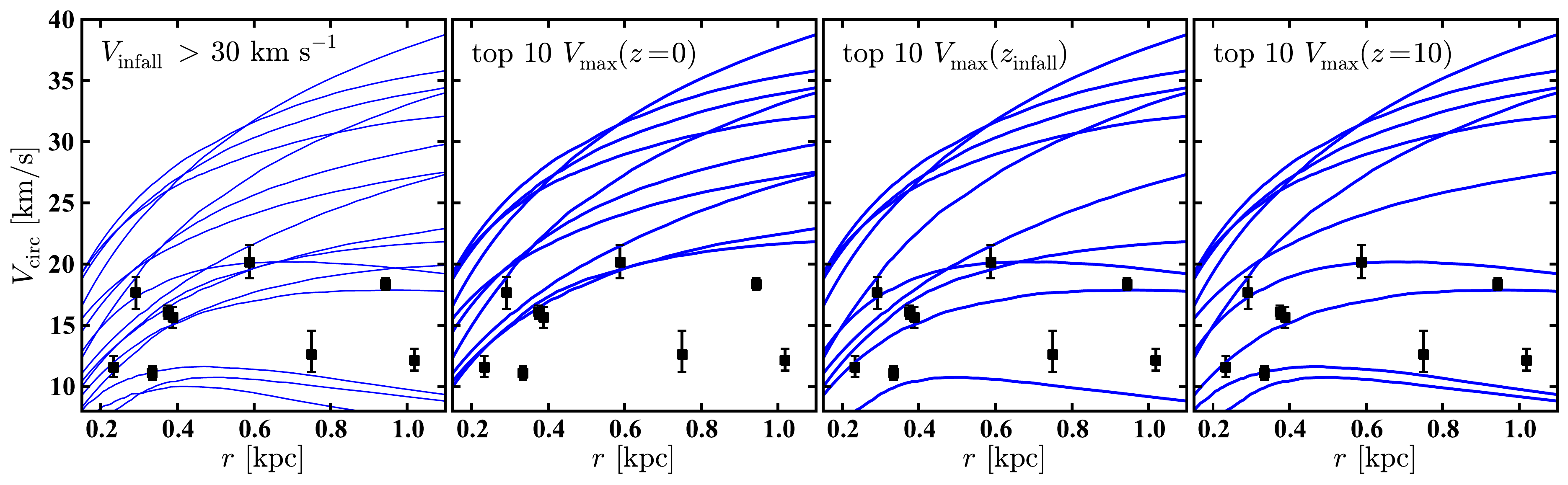}
 \includegraphics[scale=0.46]{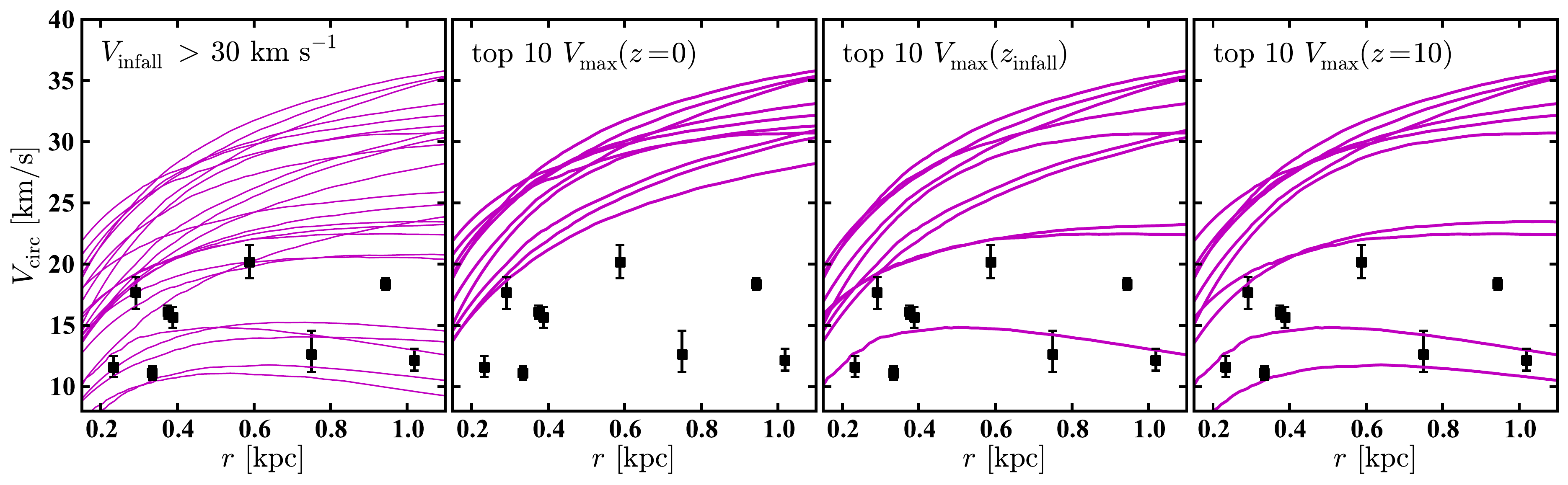}
 \caption{\textit{Left panel:} circular velocity profiles at redshift zero for
   subhalos of the Aquarius B halo (top; $\mvir = 9.5 \times 10^{11}\,\msun$)
   and E halo (bottom; $\mvir=1.4 \times 10^{12}\,\msun$) that have $\vacc >
   30\,\kms$ and $\vmax(z=0) > 10\,\kms$ (excluding MC candidates).  Measured
   $\vcirc(\rhalf)$ values for the MW dSphs are plotted as data points with
   error bars.  Each subsequent panel shows redshift zero rotation curves for
   subhalos from the left panel with the ten highest values of $\vmax(z=0)$
   (\textit{second panel}), $\vacc$ (\textit{third panel}), or $\vmax(z=10)$
   (\textit{fourth panel}).  In none of the three scenarios are the most massive
   subhalos dynamically consistent with the bright MW dSphs:
   there are always several subhalos more massive than all of the MW dSphs.
   (Analogous results are found for the other four halos.)
 \label{fig:vcirc_3ways}
}
\end{figure*}
%%%%%%%%%%%%%%%%%%%%%%%%%%%%%%%%%%%%%%%%%%%%%%%%%%%%%%%%%%%%%% 
%%%%%%%%%%%%%%%%%%%%%%%%%%%%%%%%%%%%%%%%%%%%%%%%%%%%%%%%%%%%%% 
\begin{figure*}
\centering
\includegraphics[scale=0.65]{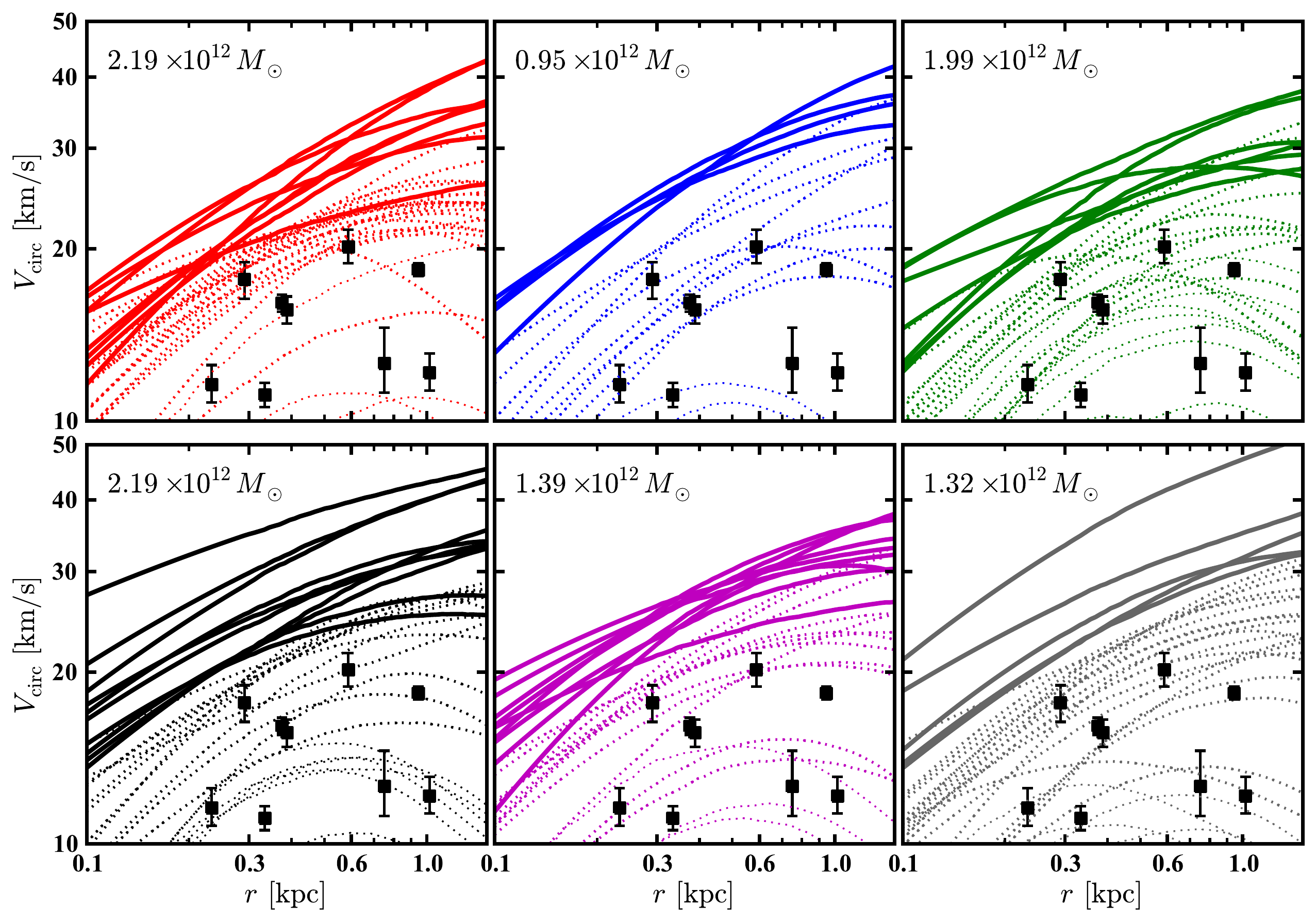}
\caption{Rotation curves for all subhalos with $\vacc >
  30\,\kms$ and $\vmax > 10 \,\kms$, after excluding Magellanic Cloud analogs,
  in each of the six Aquarius simulations (top row, from left to right: A, B, C;
  bottom row: D, E, F).  Subhalos that are at least $2\,\sigma$ denser than
  every bright MW dwarf spheroidal are plotted with solid curves, while the
  remaining subhalos are plotted as dotted curves.  Data points with errors show
  measured $\vcirc$ values for the bright MW dSphs.  Not only does each halo
  have several subhalos that are too dense to host any of the dSphs, each halo
  also has several massive subhalos (nominally capable of forming stars) with
  $\vcirc$ comparable to the MW dSphs that have no 
  bright counterpart in the MW.  In total, between 7 and 22 of these massive
  subhalos are unaccounted for in each halo.
   \label{fig:vcirc_new}
 }
\end{figure*}
%%%%%%%%%%%%%%%%%%%%%%%%%%%%%%%%%%%%%%%%%%%%%%%%%%%%%%%%%%%%%% 

The analysis in Sec.~\ref{subsec:preliminary_comparison}, based on the
assumption that subhalos obey NFW profiles, is similar to the analysis presented
in \bkbk.  On a case-by-case basis, however, it is possible that subhalos may
deviate noticeably from NFW profiles.  Consequently, \textit{the remainder of
  our analysis is based on properties of subhalos computed directly from the raw
  particle data}.  We employ a correction that takes into account the unphysical
modification of the density structure of simulated subhalos due to force
softening; this procedure is detailed in Appendix~\ref{sec:appendix}.  We note,
however, that our results do not change qualitatively if we neglect the
softening correction (see Appendix~\ref{sec:appendix} and
Table~\ref{table:posteriors_append}).  By using the particle data directly, we
remove any uncertainties originating from assumptions about the shape of the
subhalos' density profiles.
 
The consistency between massive \lcdm\ subhalos and the bright dSphs of the MW
is assessed in Figure~\ref{fig:vcirc_3ways}.  As there is strong theoretical
motivation to believe it is $\vacc$ rather than $\vmax(z=0)$ that correlates
with galaxy luminosity, we focus on the most massive subhalos in terms of
$\vacc$ -- those with $\vacc>30\,\kms$.  We remove from this group all subhalos
that are Magellanic Cloud analogs according to the criteria given at the end of
Sec.~\ref{subsec:observations}.  The left-hand panels of the figure show circular
velocity profiles of the remaining massive subhalos in two of the Aquarius
halos, Aq-B (upper panels; $\mvir =9.5\times10^{11}\,\msun$, the lowest of the
Aquarius suite) and Aq-E (lower panels; $\mvir = 1.39 \times 10^{12}\,\msun$).
Subsequent panels show the ten most massive of these subhalos as measured at
$z=0$ (second column), $z=\zacc$ (third column), and $z=10$ (forth column).

The most massive subhalos in terms of $\vacc$ span a range of profiles at $z=0$,
as the left panel of Fig.~\ref{fig:vcirc_3ways} shows.  For each halo, some of
these massive subhalos are consistent with the observed data while others are
not.  Focusing on the most massive subhalos at the present day (second panels
from left), we see that these halos are markedly inconsistent with the dSphs,
re-enforcing the results of Sec.~\ref{subsec:preliminary_comparison}.  However,
most subhalos that are massive at $z=0$ were also massive in the past, a point
that is emphasized in the two right panels of the figure: the bright MW dSphs
are also inconsistent with either the most massive subhalos in terms of $\vacc$
or those defined by their mass at $z=10$ (a possible proxy for the mass at
reionization).  Even for Aq-B, the lowest mass host halo in the sample, four of
the ten most massive subhalos are more massive than any of the dSphs,
independent of the definition of subhalo mass.

The agreement between MW dSphs and massive subhalos is even worse for the other
five Aquarius halos.  In Fig.~\ref{fig:vcirc_new}, we compare the redshift zero
rotation curves of subhalos from each of the six Aquarius halos to the observed
values of $\vcirc(\rhalf)$ for the bright Milky Way dwarf spheroidals.  As in
Fig.~\ref{fig:vcirc_3ways}, we plot only halos with $\vacc>30\,\kms$ and
$\vmax(z=0) > 10\,\kms$.  Subhalos that are at least $2\sigma$ more massive than
\textit{every} dwarf (at $\rhalf$) are plotted as solid curves; these are the
``massive failures'' discussed in \bkbk, and each halo has at least four such
subhalos.  Fig.~\ref{fig:vcirc_new} shows that each halo has several other
subhalos with $\vacc > 30$ that are unaccounted for as well: for example, halo B
has three subhalos that are not massive failures by our definition but that are
inconsistent at $2\,\sigma$ with every dwarf except Draco.  Even ignoring the
subhalos that are completely unaccounted for (and are yet more massive than all
of the MW dSphs), the remaining massive subhalos do not resemble the bright MW
dSph population.

\subsection{High redshift progenitors of massive subhalos}
\label{subsec:high_z}
%%%%%%%%%%%%%%%%%%%%%%%%%%%%%%%%%%%%%%%%%%%%%%%%%%%%%%%%%%%%%%
\begin{figure}
 \centering
 \includegraphics[scale=0.58]{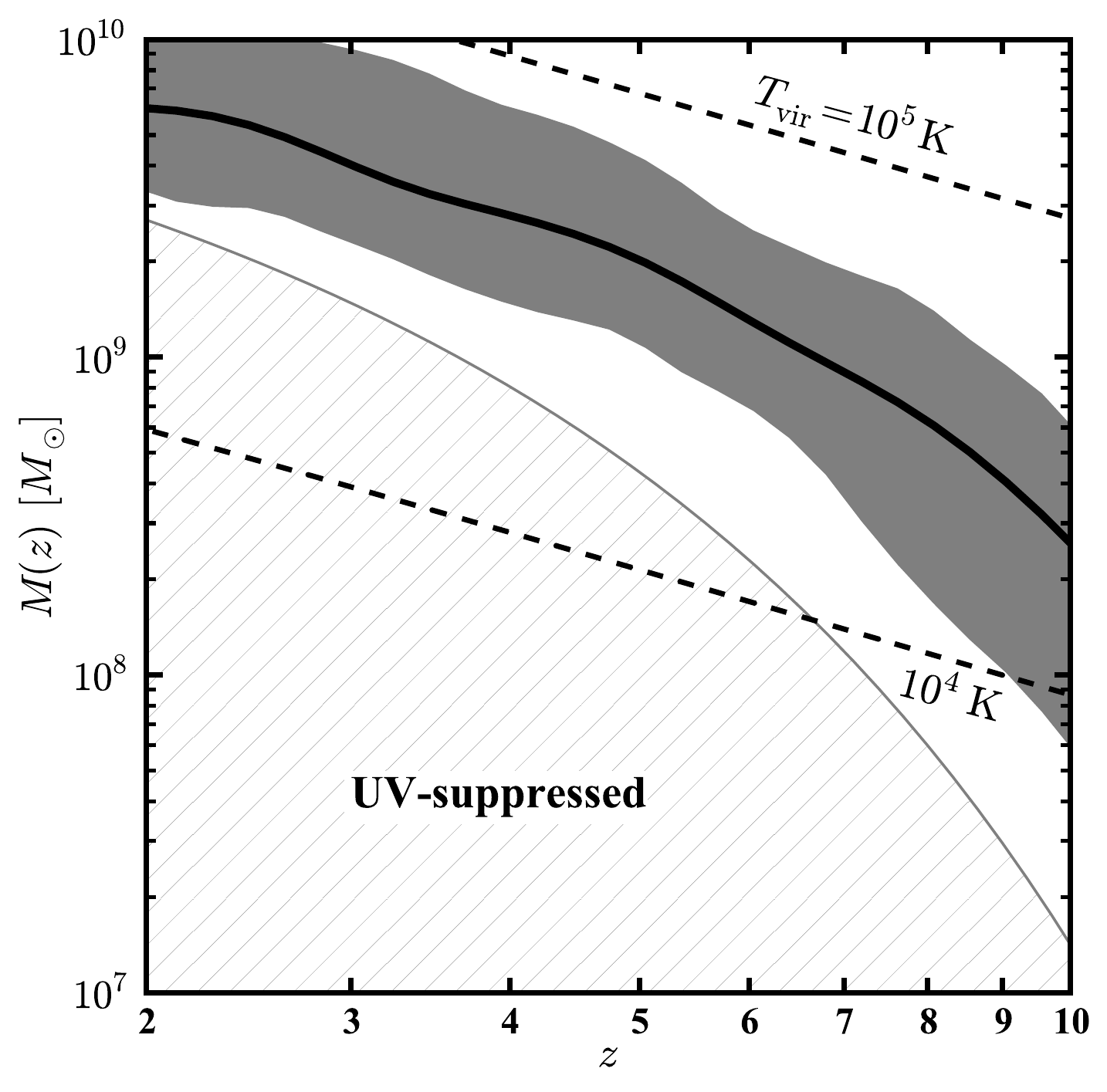}
 \caption{The median mass of $z=0$ subhalos having $\vacc > 30 \,\kms$
   (excluding MC analogs) as function of redshift (solid curve), along with the
   68\% confidence range, symmetric about the median (shaded region).  
   The hatched region marked ``UV-suppressed'' shows where halos are
   expected to have lost at least 50\% of their baryonic mass owing to the UV
   background \citep{okamoto2008}. Subhalos at $z=0$ having $\vacc > 30 \,\kms$
   are more massive than the photo-suppression scale at all redshifts.
\label{fig:mfilter_30}
}
\end{figure}
%%%%%%%%%%%%%%%%%%%%%%%%%%%%%%%%%%%%%%%%%%%%%%%%%%%%%%%%%%%%%% 
%%%%%%%%%%%%%%%%%%%%%%%%%%%%%%%%%%%%%%%%%%%%%%%%%%%%%%%%%%%%%%
\begin{figure}
 \centering
 \includegraphics[scale=0.58]{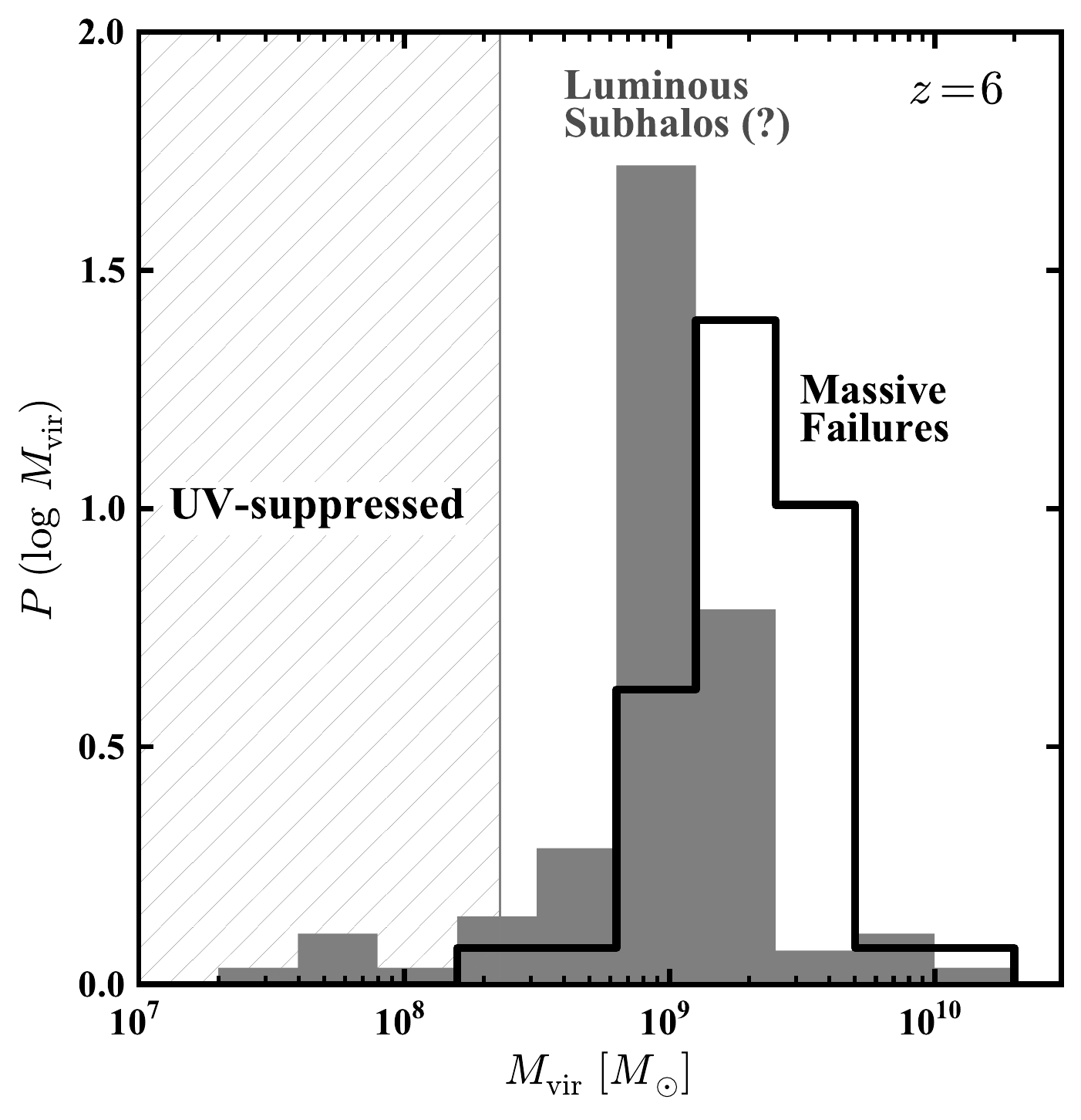}
 \caption{The distribution of masses at $z=6$ for $z=0$ subhalos with $\vacc > 
   30\,\kms$ (excluding MC candidates).  The open black histogram shows the
   ``massive failures'' (subhalos that are $2\,\sigma$ more dense than all of
   the MW's bright dSphs), while the filled gray histogram shows the remaining
   subhalos, which we deem to be potentially luminous satellites.  Even at
   $z=6$, the massive failures are typically more massive 
   than subhalos consistent with the bright dSphs.  They are also all more
   massive than 
   characteristic scale below which the UV background significantly affects the
   baryon content of halos (hatched region).   At $z=6$, this characteristic
   mass is comparable to $\tvir=10^{4}\,{\rm K}$, the threshold for atomic
   cooling. 
 \label{fig:massfunc_z6}
}
\end{figure}
%%%%%%%%%%%%%%%%%%%%%%%%%%%%%%%%%%%%%%%%%%%%%%%%%%%%%%%%%%%%%% 

To investigate the possible impact of reionization on our results, we show the
evolution of the progenitors of all subhalos with $\vacc> 30\,\kms$ in
Figure~\ref{fig:mfilter_30}.  The solid curve show the median $M(z)$, while the
shaded region contains 68\% of the distribution, centered on the median, at each
redshift.  For comparison, we also show $\tvir(z)=10^{4}\,{\rm K}$ (the
temperature at which primordial gas can cool via atomic transitions) and
$10^{5}\,\mathrm{K}$ (dashed lines), as well as the mass $M_c(z)$ below which at
least half of a halo's baryons have been removed by photo-heating from the UV
background \citep{okamoto2008}.  Subhalos with $\vacc > 30\,\kms$ lie above
$M_c$ and $\tvir=10^4\,\mathrm{K}$ at all redshifts plotted, indicating that
they are too massive for photo-ionization feedback to significantly alter their
gas content and thereby inhibit galaxy formation.  

Figure~\ref{fig:massfunc_z6} focuses on the $z=6$ properties of these subhalos.
It shows the distribution of halo masses at $z=6$ for ``massive failures'' (open
histogram) and the remaining subhalos (filled histogram), which are possible
hosts of the MW dSphs.  The massive failures are more massive at $z=6$, on
average, than the potentially luminous subhalos.  This further emphasizes that
reionization is not a plausible explanation of why the massive failures do not
have stars: the typical massive failure is a factor of ten more massive than the
UV suppression threshold at $z=6$.  Implications of this result will be
discussed in Boylan-Kolchin et al. (in preparation).

In a series of recent papers, Broderick, Chang, and Pfrommer
\nocite{broderick2011, chang2011, pfrommer2011} have postulated that the thermal
history of the IGM at late times ($z \la 2-3$) could differ substantially from
standard reionization models owing to a large contribution from TeV blazars.
This modification relies on plasma instabilities dissipating energy from TeV
blazars in the IGM, heating it to a temperature that is a factor of $\sim \!
3-10$ higher than in the case of pure photo-ionization heating.  Such heating
would effectively increase the value of $M_c(z)$ for $z \la 2-3$, suppressing
the stellar content of more massive halos.  However, as
Figure~\ref{fig:mfilter_30} shows, all halos with $\vacc > 30\,\kms$ should have
been able to form stars before this epoch, i.e., TeV blazar heating happens too
late to kill off star formation in the Milky Way's massive subhalos (recall that
we find 16-33 subhalos with $\vacc > 30\,\kms$ per halo).  While TeV blazar
heating therefore may help reduce the counts of void galaxies (which form later
than MW subhalos) and suppress the star formation at low redshifts in
progenitors of MW subhalos, it does not seem capable of explaining the structure
and abundance of massive MW satellites.

\section{Dark matter masses of the Milky Way dwarfs}
\label{sec:dwarf_dm_mass}
The results of Sec.~\ref{sec:lcdm_mw} show that the brightest Milky Way dwarfs do not
inhabit the most massive dark matter subhalos from numerical simulations.  We
can also use the simulations to compute the properties of subhalos that are
consistent with the dynamics of the bright dSphs.  These calculations, and the
resulting implications, are the subject of this Section.  

\subsection{Computing properties of the dark matter hosts of the dwarf
  spheroidals }
\label{subsec:posteriors}
To compute more rigorous estimates of properties of the dSphs' dark matter
halos, we assume that the subhalo population across all six Aquarius halos forms
a representative sample from \lcdm\ simulations.  We can then compute the
distribution function of $X$ (where $X$ is, for example, $\vmax, \, \vacc$, or
$\macc$) for a dwarf by assigning a weight (likelihood $\mathcal{L}$) to each
subhalo in our sample based on how closely it matches the measured $\mhalf$
value of that dwarf.  The posterior distribution of quantity $X$, given the
measured value of $\mhalf$ and its error $\sigma_M$, is given by
\begin{equation}
  \label{eq:1}
  {\rm P}(X | \mhalf, \sigma_{M}) \propto {\rm P}(X) \, 
  \mathcal{L}(X|\mhalf, \sigma_{M})\,.
\end{equation}

In practice, we compute moments in the distribution of quantity $X$ via
\begin{equation}
  \label{eq:2}
    \langle X^\alpha \rangle =  \frac{\displaystyle\sum_{i=1}^{N_{\rm subs}}
      X_i^{^\alpha} \,{\rm P}(X_i)\,\mathcal{L}(X_i|\mhalf, \sigma_{M})}
    {\displaystyle\sum_{i=1}^{N_{\rm subs}}{\rm P}(X_i)\,\mathcal{L}(X_i |\mhalf, \sigma_{M})} \,
\end{equation}
to compute properties of the hosts of the MW dwarf spheroidals.  We assume that
the likelihood functions are log-normal, and compute the relevant moments in
$\log$ as well, e.g., $\langle \log \vmax \rangle$ and $\sigma^2_{\log \vmax}
\equiv \langle (\log \vmax)^2-\langle \log \vmax \rangle^2 \rangle$.  

\begin{table}
  \caption{
    Derived masses of the bright Milky Way dSphs. \textit{Columns}: (1) Name of
    dwarf; (2) $V$-band luminosity (from \citealt{wolf2010}); (3) median
    $\vmax$; (4) median $\vacc$; (5) median $\macc$.  The quantities in columns
    3-5  are computed using Eqn.~\eqref{eq:2}; the errors represent the 68.3\%
    confidence interval. 
 \label{table:posteriors}
  }
 \begin{tabularx}{\columnwidth}{@{\extracolsep{\fill}} lccccc}
  \hline
  \hline
  Name & $L_V$ & $\vmax$ & $\vacc$ & $\macc$\\
  & [$\lsun$] \Bspace & [$\kms$] & [$\kms$] & [$\msun$]\\
 \hline
Fornax \Tspace \Bspace & $1.7^{+0.5}_{-0.4} \times 10^{7}$ & $17.8^{+0.7}_{-0.7}$ &  $22.0^{+4.7}_{-3.9}$ & $7.4^{+6.1}_{-3.3} \times 10^{8}$ \\
LeoI \Tspace \Bspace & $5.0^{+1.8}_{-1.3} \times 10^{6}$ & $16.4^{+2.3}_{-2.0}$ &  $20.6^{+5.7}_{-4.5}$ & $5.6^{+6.8}_{-3.1} \times 10^{8}$ \\
Sculpt \Tspace \Bspace & $2.5^{+0.9}_{-0.7} \times 10^{6}$ & $17.3^{+2.2}_{-2.0}$ &  $21.7^{+5.8}_{-4.6}$ & $6.6^{+7.8}_{-3.6} \times 10^{8}$ \\
LeoII \Tspace \Bspace & $7.8^{+2.5}_{-1.9} \times 10^{5}$ & $12.8^{+2.2}_{-1.9}$ &  $16.0^{+4.7}_{-3.6}$ & $2.4^{+3.1}_{-1.4} \times 10^{8}$ \\
Sextans \Tspace \Bspace & $5.9^{+2.0}_{-1.4} \times 10^{5}$ & $11.8^{+1.0}_{-0.9}$ &  $14.2^{+3.7}_{-2.9}$ & $1.9^{+1.7}_{-0.9} \times 10^{8}$ \\
Carina \Tspace \Bspace & $4.3^{+1.1}_{-0.9} \times 10^{5}$ & $11.4^{+1.1}_{-1.0}$ &  $14.4^{+3.7}_{-3.0}$ & $1.8^{+1.8}_{-0.9} \times 10^{8}$ \\
UrsMin \Tspace \Bspace & $3.9^{+1.7}_{-1.3} \times 10^{5}$ & $20.0^{+2.4}_{-2.2}$ &  $25.5^{+7.4}_{-5.8}$ & $1.1^{+1.5}_{-0.6} \times 10^{9}$ \\
CVnI \Tspace \Bspace & $2.3^{+0.4}_{-0.3} \times 10^{5}$ & $11.8^{+1.3}_{-1.2}$ &  $14.5^{+4.0}_{-3.1}$ & $1.9^{+2.0}_{-1.0} \times 10^{8}$ \\
Draco \Tspace \Bspace  & $2.2^{+0.7}_{-0.6} \times 10^{5}$ & $20.5^{+4.8}_{-3.9}$ &  $25.9^{+8.8}_{-6.6}$ & $1.2^{+2.0}_{-0.7} \times 10^{9}$ \\
\hline
\end{tabularx}
\end{table}

The resulting values, with $1 \,\sigma$ errors, for $\vmax,\,\vacc$, and $\macc$
are listed in Table~\ref{table:posteriors}.  This analysis confirms the
preliminary comparison performed in Section~\ref{subsec:preliminary_comparison}.
The dwarfs are all consistent with subhalos in the range of $10 \la \vmax \la 25
\,\kms$, and at 95\% confidence, {\it none} of the nine dwarfs have halos of
$\vmax =30 \,\kms$ or greater.  The central values of $\vacc$ range from $\sim
14-26\,\kms$, with all but Draco and Ursa Minor having $\vacc < 35\,\kms$
($2\,\sigma$).  Each simulation has 12-22 (16-33) subhalos with $\vacc >
35\,(30)\,\kms$ and $\vmax > 10\,\kms$, however (the same is true for the Via
Lactea II simulation: we find 27 subhalos with both $\vmax > 10\,\kms$ and
$\vacc > 30\,\kms$).  Even including for the Magellanic Clouds and Sagittarius,
there are at least 6-21 subhalos in each simulation that are unaccounted for but
have high enough masses that they should be luminous.  Furthermore, the
satellites predicted to be hosted by the most massive subhalos, Draco and Ursa
Minor, are two of the three least luminous satellites in our sample.

Eq.~\eqref{eq:2} assumes that the probability of a subhalo hosting a specific
dwarf depends only on how well that subhalo's $M(r_{\rm 1/2, dwarf})$ agrees
with the measured $\mhalf$ of the dwarf.  If subhalos spanning a range of
$\vmax$ or $\vacc$ have identical values of $M(r_{\rm 1/2, dwarf})$,
Eq.~\eqref{eq:2} assigns each of them the same probability of hosting that
dwarf.  In this case, the resulting mean value of $\vmax$ may get larger weight
from the numerous low-mass subhalos than from the less abundant high-mass
subhalos since the mass function of CDM subhalos scales as $N(>V) \propto
V^{-3}$.  We have repeated our analysis with an additional weighting factor of
$V^3$ to mimic a prior of equal probability per unit $\log \, V$ (effectively
giving massive subhalos higher weights), and find that our results are
qualitatively unchanged.  This is not surprising, as the data strongly constrain
$\mhalf$, which, for subhalos, is tightly correlated with $\vmax$: the results
are driven by the data, not by the choice of prior.

\citet{strigari2007}, \citet{madau2008a}, and \citet{kuhlen2010} have previously
used the Via Lactea simulations to derive constraints on $\vmax$ (and, in the
case of \citealt{kuhlen2010}, $\vacc$) values.  These were based on the masses
of the satellites within either 300 or 600 pc ($\mthree$ or $\msix$).  Our
calculations are based on a larger sample of \lcdm\ subhalos and use more
recently determined dynamical constraints on the dwarfs -- the masses with their
de-projected half light radii -- that have smaller errors than do $\mthree$ or
$\msix$.  Furthermore, we have attempted to correct for the numerical effect of
gravitational softening that affects $\mthree$ at the $\sim\!  20\%$ level in
simulations with the force resolution of Aquarius level 2.  This results in a
decrease in the derived $\vmax$ values relative to the uncorrected case: halos
of a fixed $\vmax$ have larger $\mthree$ values after applying the correction.
While our approach is somewhat more detailed than that of \citet{madau2008a} and
\citet{kuhlen2010}, our results are generally consistent with those studies, and
with \citet{strigari2007}.  Our results are also consistent with those of
\citet*{strigari2010}, who have tested whether photometric and kinematic data on
five of the bright dSphs are in accord with the gravitational potentials of
Aquarius subhalos.  They found good matches in each of the five cases, but
always in systems with $\vmax$ values of $10-30\,\kms$, never in more massive
halos.  We find somewhat smaller $\vmax$ values for Draco, Leo I, Ursa Minor,
and Sculptor than \citet{penarrubia2008}, likely because they adopted the
$\vmax-\rmax$ relation for field dark matter halos: \citet{springel2008} find
that the Aquarius subhalos are systematically denser than field halos, with an
offset of 0.2 dex in $\rmax$ at fixed $\vmax$.

\subsection{Comparison to $\bmlcdm$ predictions}
\label{subsec:cdm_compare}
%%%%%%%%%%%%%%%%%%%%%%%%%%%%%%%%%%%%%%%%%%%%%%%%%%%%%%%%%%%%%%
\begin{figure*}
 \centering
 \subfloat{\includegraphics[scale=0.55]{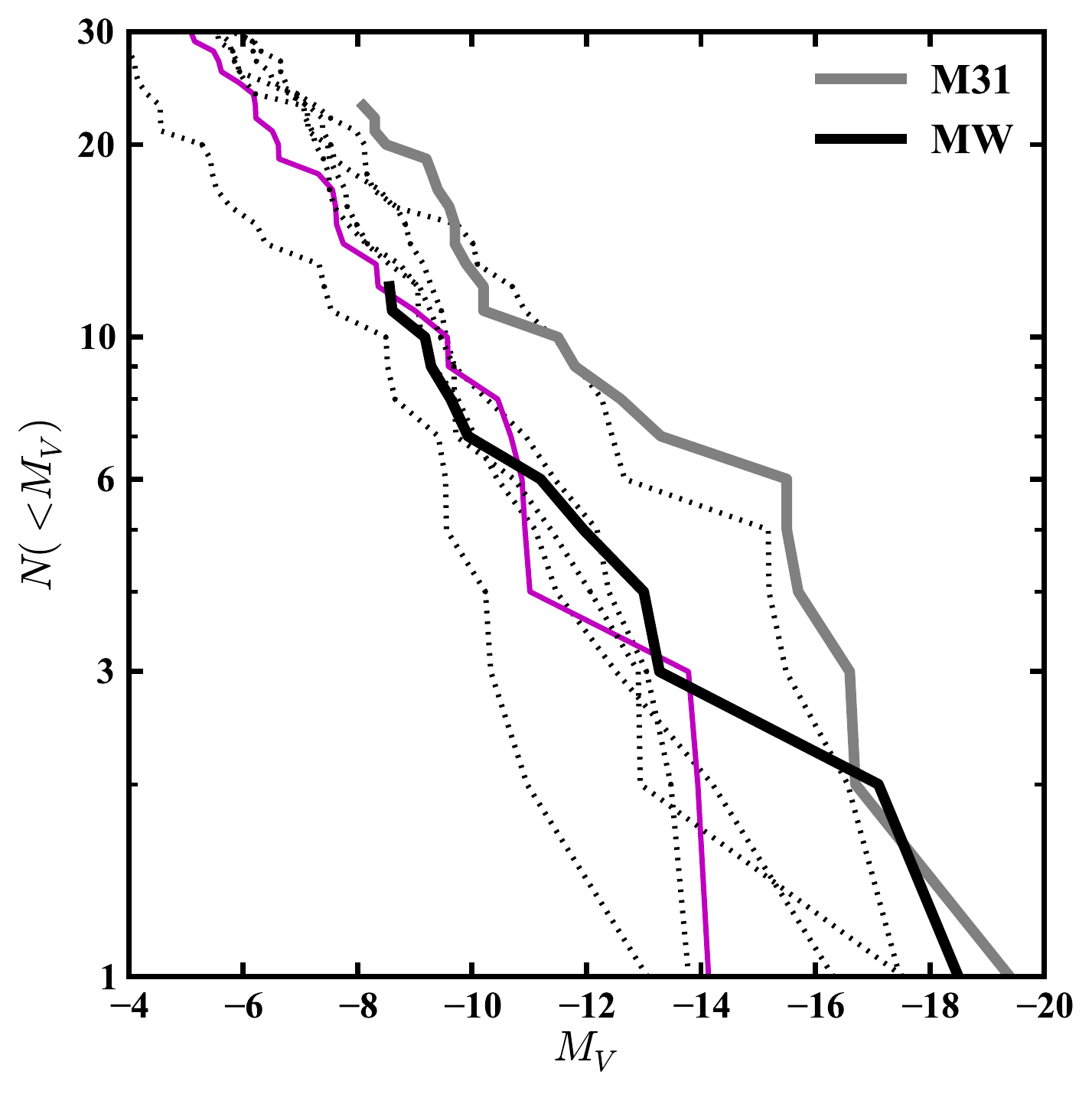}}
 \subfloat{\includegraphics[scale=0.55]{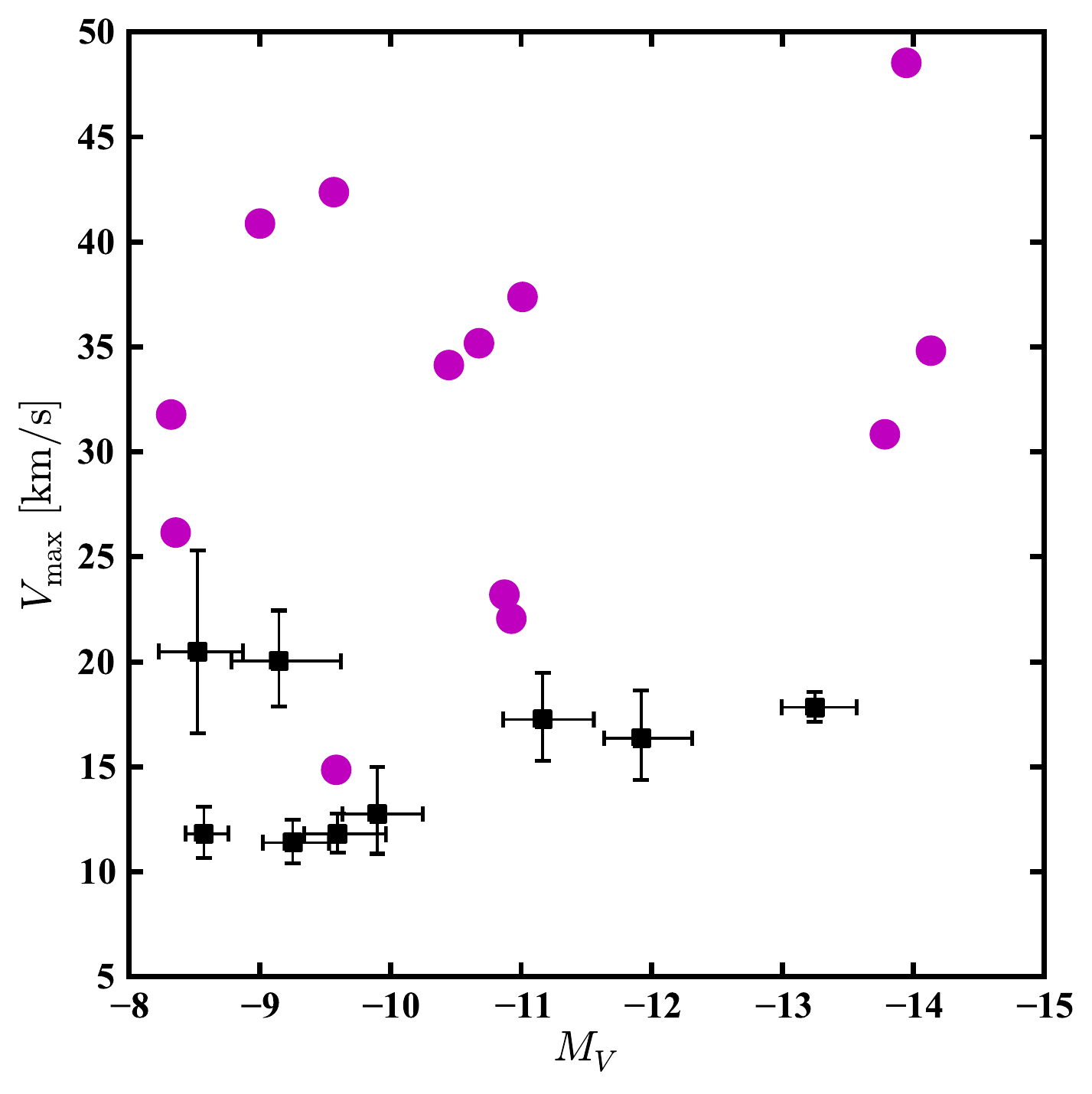}}
 \caption{\textit{Left}: Observed luminosity functions for the Milky Way and M31
   (thick solid lines) compared to abundance matching predictions based on the
   Aquarius simulations (thin lines, with Aq-E plotted in magenta;
   $\mstar/L_V=2$ is assumed).  \textit{Right}: Values of $\vmax$ computed in
   Sec.~\ref{subsec:posteriors} for the nine luminous Milky Way dwarf
   spheroidals (square symbols with errors), along with $\vmax(z=0)$ values of
   the subhalos with $M_V < -8$ (magnitudes are assigned by abundance matching)
   from the halo that best reproduces the luminosity function in the
     left panel (Aq-E).  While numerical simulations combined with
   abundance matching reproduces the luminosity function of MW satellites, the
   structure of the dwarf spheroidals hosts' in this model does not match
   observations: the simulated subhalos are much more massive (have larger
   values of $\vmax$) than the dSphs.
   \label{fig:sham_lf}
 }
\end{figure*}
%%%%%%%%%%%%%%%%%%%%%%%%%%%%%%%%%%%%%%%%%%%%%%%%%%%%%%%%%%%%%% 
We are now in a position to directly compare observations and theoretical
predictions for the hosts of the MW dSphs.  The left-hand panel of
Fig.~\ref{fig:sham_lf} shows the luminosity functions of the Milky Way and M31
(solid black and gray curves) compared to the predicted luminosity functions
from the Aquarius simulations using abundance matching, with $\mstar/L_V=2$
(dotted curves).  The abundance matching relation itself is computed by equating
subhalo abundances from the Millennium and Millennium-II Simulations (closely
following \citealt{guo2010}) to galaxy abundances from the SDSS \citep{li2009},
with a power-law extrapolation to lower halo masses.  While few of the simulated
halos have subhalos massive enough to host the MW's brightest satellites, the
Magellanic Clouds \citep{boylan-kolchin2010, boylan-kolchin2011, busha2011},
the agreement on the dSph scale ($M_V > -14$) is remarkably good in 5 of the 7
halos.  Note that this agreement is \textit{not} built into the abundance
matching model: it results from the MW satellite luminosity function having both
the same slope and amplitude as the extrapolation of the field galaxy luminosity
function to (much) lower luminosities than can currently be probed
statistically, which is non-trivial.

The masses of the subhalos that abundance matching predicts should host the
dSphs are very different from those of the observed dSphs, however; this
important point is illustrated in the right-hand panel of
Fig.~\ref{fig:sham_lf}.  The black squares with errors show the $\vmax$ values
of the MW dSphs derived in Sec.~\ref{subsec:posteriors}, while the magenta
circles show the measured $\vmax$ values from one of the numerical simulations
that matches the luminosity function well (also colored magenta in the left
panel of Fig.~\ref{fig:sham_lf}).  The $\vmax$ values of the simulated subhalos
are systematically higher than those of the MW dSphs.  \textit{It is therefore
  not possible to simultaneously match the \emph{abundance} and \emph{structure}
  of the MW dSphs in standard galaxy formation models based on dissipationless
  \lcdm\ simulations}.  While there are many subhalos that match the structure
of the bright MW dSphs, these are \textit{not} the subhalos that are predicted
to host such galaxies in \lcdm.  \vspace{0.5cm}

%%%%%%%%%%%%%%%%%%%%%%%%%%%%%%%%%%%%%%%%%%%%%%%%%%%%%%%%%%%%%%
\begin{figure}
  \centering
  \subfloat{\includegraphics[scale=0.56]{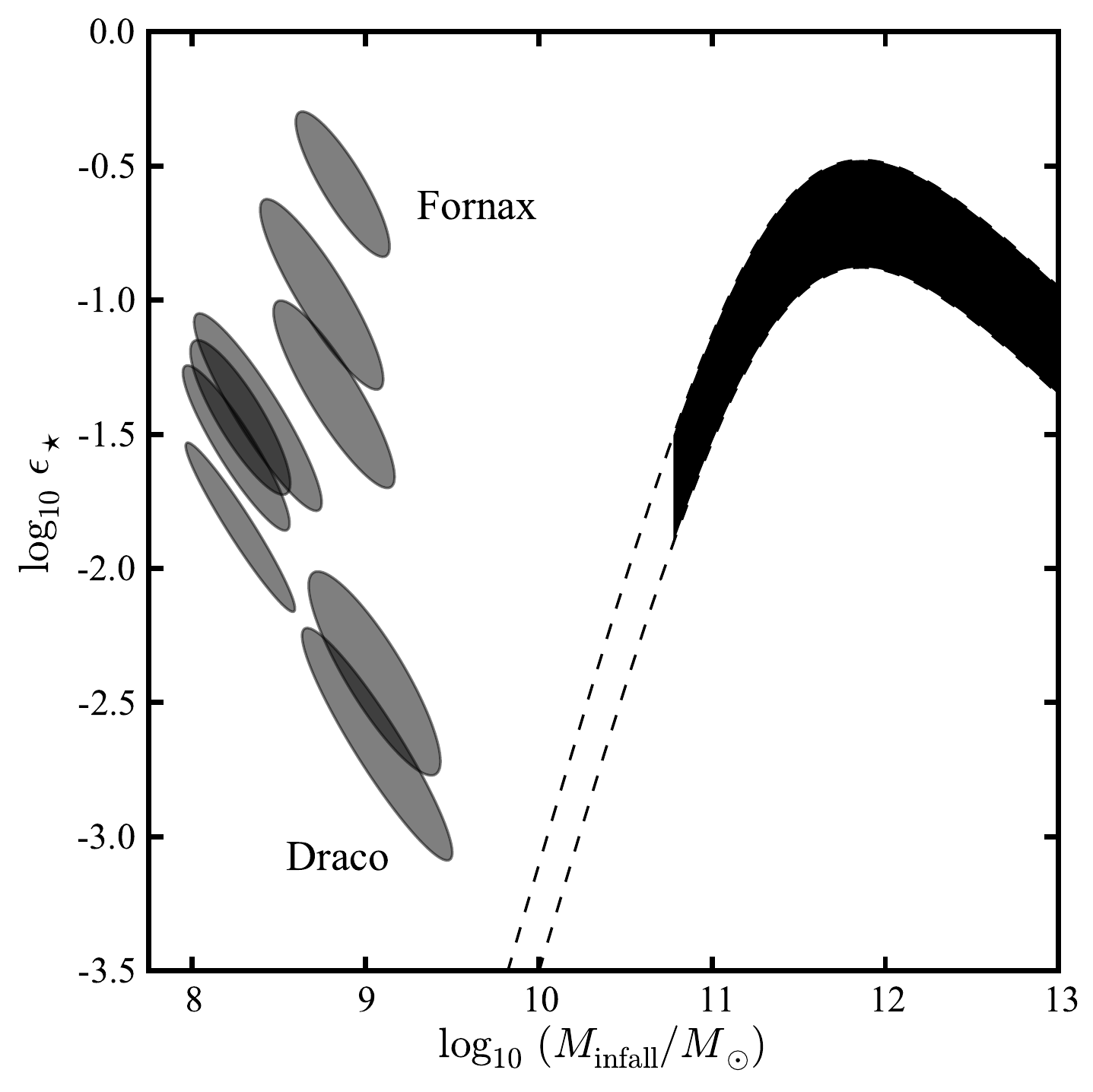}}
  \caption{The relation between
    $\macc$ and star formation efficiency $\estar=\mstar/(f_b\,\macc)$, the
    measure of the fraction of available baryons converted into stars in a
    halo.  Shaded ellipses show the nine bright dSphs
    considered in this paper, encompassing the $1\,\sigma$ uncertainties in
    $\macc$ from Table~\ref{table:posteriors}.
    The black shaded band and its extrapolation to lower
    masses (dashed lines) show $\estar(\macc)$ derived from abundance matching.
    \textit{All} of the bright dSphs have conversion efficiencies that are much
    higher than expected from abundance matching, given their masses.  Although
    the dSphs have similar values of $\macc$, their inferred conversion
    efficiencies vary by two orders of magnitude.  At larger mass scales, this
    spread is closer to a factor of two at fixed $\macc$.
    \label{fig:eff_macc}
  }
\end{figure}
%%%%%%%%%%%%%%%%%%%%%%%%%%%%%%%%%%%%%%%%%%%%%%%%%%%%%%%%%%%%%% 
The observed densities of MW satellites are very difficult to reconcile with
\lcdm-based galaxy formation models, where the stellar content of a galaxy is
strongly coupled to $\vacc$.  To highlight the problem, we plot the inferred
star formation efficiency -- $\estar \equiv \mstar / (f_b \, \macc)$, where
$f_b=\Omega_b/\Omega_m$ is the universal baryon fraction -- as a function of
$\macc$ in Fig.~\ref{fig:eff_macc}.  The ellipses show $1\,\sigma$ uncertainties
(note that the direction of the ellipses is due to the inverse correlation
between $\estar$ and $\macc$ at fixed $\mstar$). This relation is
well-constrained at $z=0$ in the context of abundance matching for $\mstar >
10^{8.3}\,\msun$ (approximately the completeness limit of the \citet*{li2009}
stellar mass function, corresponding to $\mhalo=6 \times 10^{10}\,\msun$).  The
relation for $\mstar$ lower than the SDSS completeness limit is extrapolated
using a power law (dashed portion of abundance matching lines).

The $\mstar-\mhalo$ relation cannot be tested statistically on mass scales
relevant for the dSphs at present, but it is immediately apparent that galaxy
formation must proceed differently at $\mhalo \la 10^{10}\,\msun$ than for
larger systems if simulated subhalos accurately reflect the densities of the
halos of dSphs as they exist the Universe.  For example, the most luminous dSph
of the MW, Fornax, has an inferred star formation efficiency of $\estar \approx
0.2$, a value that is approached only at the scale of MW-mass halos.  Ursa Minor
and Draco, which are $\sim 40-80$ times less luminous than Fornax, sit in halos
that are comparable or slightly more massive, and therefore have inferred
efficiencies of closer to $\estar=0.002$.

\section{Discussion}
\label{sec:discussion}

Sections~\ref{sec:lcdm_mw} and \ref{sec:dwarf_dm_mass} have demonstrated that
the structure and abundance of bright Milky Way satellites are not consistent
with populating the most massive subhalos in hosts of $\mvir \approx (1-2)\times
10^{12}\,\msun$.  In this Section, we discuss some possible remedies for this
problem, ranging from downward revisions of the MW's dark matter halo mass
(Sec.~\ref{subsec:mw_mass}) to changes to \lcdm\ (Sec.~\ref{subsec:dm_physics}).

\subsection{Mass of the Milky Way}
\label{subsec:mw_mass}
The simulated halos used in this paper range from $\mvir=9.5 \times 10^{11}$ to
$\mvir=2.2 \times 10^{12} \,\msun$.  The true mass of the Milky Way is still a
matter of significant uncertainty, however.  The apparent lack of massive
subhalos might be understandable if the Milky Way is significantly less massive
than this simulated range.  Here we summarize recent estimates of the Milky Way
halo mass.

\begin{itemize}
\item \textit{halo tracers}\\
  \citet{xue2008} used blue horizontal-branch stars from the Sloan Digital Sky
  Survey, combined with mock observations of hydrodynamical simulations of Milky
  Way-like galaxies, to find $M_{\rm vir,MW}=1.0^{+0.3}_{-0.2} \times
  10^{12}\,\msun$, and $M(<60 \,\kpc)=(4.0 \pm 0.7) \times 10^{11}\,\msun$.
  Through a Jeans analysis of halo stars obtained from a survey for
  hypervelocity stars, \citet{gnedin2010} found $M_{\rm vir,MW}=(1.6 \pm 0.3 )
  \times 10^{12}\,\msun$, and $M(<80 \,\kpc)=6.9^{+3.0}_{-1.2} \times
  10^{11}\,\msun$.  The largest uncertainties in these studies are the velocity
  anisotropy $\beta$ and density profile (slope) assumed for the halo stars.
  Both \citet{xue2008} and \citet{gnedin2010} find most likely values for
  $\beta$ that are near $0.4$, i.e., biased toward radial orbits.  A smaller
  value of $\beta$ would result in increases in the mass estimates.

\item \textit{satellite kinematics} \\
  Kinematics of satellite galaxies in the Milky Way provide constraints on its
  mass.  These constraints are also sensitive to the assumed velocity anisotropy
  of the satellite population and to whether or not Leo I, which has a large
  line-of-sight velocity, is considered a bona fide satellite.
  \citet{watkins2010} find a ``best estimate'' -- including Leo I -- of $M(<300
  \,\kpc)=(2.7 \pm 0.5) \times 10^{12}\,\msun$, which assumes that the satellite
  orbits are tangentially biased (based on observed proper motions); this
  estimate becomes $(1.4 \pm 0.3) \times 10^{12} \,\msun$ for isotropic orbits,
  and $(1.2 \pm 0.2) \times 10^{12}\,\msun$ for radially anisotropic orbits
  (consistent with the Xue et al. and Gnedin et al. results, which also assume
  such orbits).

\item \textit{dynamics of the Large Magellanic Cloud}\\
  Measurements of the proper motion of the LMC place it at a velocity of
  $\sim \! 360\, \kms$ \citep{kallivayalil2006, piatek2008}, with statistical
  errors of less than $5\%$.  This result implies that the
  Magellanic Clouds are likely on their first passage about the MW
  \citep{besla2007}, although this conclusion may depend on the mass of the MW
  \citep{shattow2009}.  Analyses of large cosmological simulations show that
  objects like the LMC are extremely unlikely to be unbound to their hosts
  \citep{boylan-kolchin2011}, which allows for estimates of the MW's virial mass
  based on the dynamics of the Magellanic Clouds.  Results of such estimates
  range from $(1.2-2.0)\times 10^{12}\,\msun$, depending on the treatment of
  baryonic physics and assumptions about the masses of the Clouds
  \citep{boylan-kolchin2011, busha2011}.

\item \textit{masers}\\
  Masers can be used to constrain the angular speed of the local standard of
  rest, which, when combined with an estimate of the distance to the Galactic
  center, give the circular velocity of the MW at the sun's location.
  \citet{reid2009} used this technique to derive $V=254\pm 16 \,\kms$,
  significantly higher than the IAU standard value of $220\,\kms$ (and therefore
  indicating a massive dark matter halo).  \citet{ando2011} also used the maser
  method but found a circular velocity of $213 \pm 5\,\kms$, in strong
  disagreement with the Reid et al. value.  This discrepancy is partially, but
  not completely, explained by different adopted distances to the Galactic
  Center $R_0$: using $R_0=8.3\,\kpc$ (as in Reid et al.), Ando et al.'s
  estimate moves to $227\,\kms$.  A combined Bayesian analysis by
  \citet{bovy2009}, which incorporates maser distances but uses greater freedom
  in the modeling than was allowed for in either Reid et al. or Ando et
  al., gives $V_c(R_0)=236 \pm 11\,\kms$.  This is higher than, but marginally
  consistent with, the IAU value.  
  
  Converting between $V_c(R_0)$ and $\mvir$ is non-trivial.  We can take some
  guidance from the MW mass models of \citet*{klypin2002}, however.  They found
  that the lowest halo mass that reasonably fit the MW was $\mvir = 7.1\times
  10^{11}\,\msun$, which gave $V_c(R_0)=216 \,\kms$ (if angular momentum
  exchange between baryons and dark matter is
  included) or $246 \,\kms$ (without angular momentum exchange).  These values
  require a maximal disk of $\mstar=6\times 10^{10}\,\msun$, and a value of
  $M(<\!100\,\kpc)=3.8\times 10^{11}\,\msun$ that is much lower than the more
  recent values of \citet{gnedin2010} and \citet{mcmillan2011}.  The favored
  models of \citet{klypin2002} assume $\mvir =1.0 \times 10^{12}\,\msun$ and
  $\vmax \approx 220-230\,\kms$.

\item \textit{other measures}\\
  \citet{mcmillan2011} adopted a Bayesian approach to constraining the mass of
  MW, incorporating photometric and kinematic data as well as theoretical
  expectations from \lcdm\ and modeling of the Galaxy.  The result was a
  best-fitting mass within 100 kpc of $M(<100\,\kpc)=(0.84 \pm 0.09) \times
  10^{12} \msun$, and a virial mass of $\mvir=(1.26 \pm 0.24) \times
  10^{12}\,\msun$.  Based on abundance matching, \citet{guo2010} found $M_{\rm
    MW}= 1.99\times 10^{12}\,\msun$, with an 80\% confidence interval of
  $(0.8-4.74) \times 10^{12}\,\msun$.  (These numbers are based on an
  overdensity criterion of $\Delta = 200$; using our value of $\Delta = 94$
  moves this range to $(0.95-5.7) \times 10^{12}\,\msun$.)
\end{itemize}

The consensus of these results, then, is that the MW has a virial mass of $\sim
(1.0-2.0) \times 10^{12}\,\msun$.  The Aquarius halos used in this paper are
broadly consistent with this range.  Three of the halos -- A, C, and D -- lie at or
slightly above $2 \times 10^{12} \,\msun$, while one -- B -- lies just below
$10^{12}\,\msun$.  That all of these halos have a substantial population of
subhalos that are unaccounted for by the bright dSphs is a strong sign that the
existence of ``massive failures'' is not an artifact of the choice of halo
masses for the Aquarius halos.  Nevertheless, it is important to consider the
possibility that the MW's halo is less massive than is currently indicated by
data.  

The substructure abundance in within the virial radii of dark matter halos is
essentially self-similar when measured in terms of $\vmax/\vvir$ (e.g.,
\citealt{gao2004b, de-lucia2004, van-den-bosch2005, gao2011}), and the
cumulative abundance of dark matter subhalos in a given halo scales as
$N(>\!\vmax) \propto \vmax^{-3}$ \citep{springel2008}, so the number of halos of
fixed $\vmax$ will scale linearly with $\mvir$.  If the Aquarius halos are more
massive than the halo of the Milky Way, then our first-order expectation is that
the number of subhalos that are unaccounted for by the bright MW satellites will
decrease by an amount proportional to the necessary reduction in virial mass.

Since we find at least 6 subhalos with $\vacc > 30 \,\kms$ that are unaccounted
for when $\mvir=9.5\times 10^{11}\,\msun$ and 16-21 such subhalos when
$\mvir=2.2\times 10^{12} \msun$, a reasonable scaling seems to be $N_{\rm extra}
=6 \, (\mvir/10^{12}\,\msun)$.  Reducing the halo mass to $5 \times
10^{11}\,\msun$ would then result in $N_{\rm extra} \approx 3$, which perhaps
could be explained by halo-to-halo fluctuations.  However, for the Milky Way to
have a virial mass of $5 \times 10^{11}\,\msun$, all of the following would need
to be true: (1) the LMC, SMC, and Leo I are all unbound to the Milky Way; (2)
the Magellanic Clouds are \textit{extreme} outliers in terms of their (high)
masses; (3) stars in the Galactic halo are have high (radially biased) velocity
anisotropies; (4) either the Local Group has a mass that is substantially
smaller than what is derived based on the timing argument \citep{li2008a}, or
M31 is 3-4 times more massive than the Milky Way in spite of having a fairly
similar rotation velocity.  

Even if this is the case, it would still not explain the gap in $\vmax$ between
the most massive dwarf spheroidals ($\la 25\,\kms$; Fig.~\ref{fig:sham_lf},
right-hand panel) and Magellanic Clouds ($\vcirc \ga 50 \,\kms$): \lcdm\
simulations do not contain such a gap.  We therefore believe that a downward
revision of the MW's mass is not the most likely explanation of the massive
failures.  As noted in Sec.~\ref{subsec:simulations}, the abundance and
structure of subhalos in the VL-II simulation (the only other simulation that
has published data resolving subhalos on scales relevant to structure of the MW
dwarfs) is very similar to that of the Aquarius simulations, providing no
evidence that the background cosmology of the simulations causes the
discrepancy.

Finally, in the unlikely scenario where the Milky Way's mass has been
systematically over-estimated, the unexpected span in implied star
formation efficiencies demonstrated in Fig.~\ref{fig:eff_macc} would still
persist.  The only way to address the massive failures problem
(Figs.~\ref{fig:vcirc_3ways} -- \ref{fig:sham_lf}) and the efficiency problem
(Fig.~\ref{fig:eff_macc}) simultaneously is to modify the inner densities of
massive dark matter subhalos significantly, and in a way that is anti-correlated
with stellar mass.  In the next section we explore whether baryonic feedback can
plausibly remedy the situation.

\subsection{Baryonic feedback?}
\label{subsec:feedback}
Energy or momentum-driven feedback from star formation is expected to drive
large-scale outflows in galaxies at a variety of masses and redshifts.  Although
this feedback cannot couple hydrodynamically to dark matter at the centers of
the halos of the Milky Way dwarfs, the removal of substantial amounts of gas
would cause a dynamical re-arrangement, thereby reducing the central density of
dark matter.  The problems discussed above may be remedied if the central
regions of the rotation curves of massive subhalos can be sufficiently reduced
by such blow-out.  Here, we aim to estimate the magnitude of this effect due to
kinetic outflows.

A supernova will release $\sim 10^{51}$ ergs of energy into its surroundings, a
fraction $\epsilon_{\rm SN}$ of which will couple to the gas.  If we parametrize
the number of supernova explosions per 100 solar masses of stars formed as
$N_{100}$, then the gas mass carried by a supernova-driven outflow (with
velocity $V_{\rm out}$) due to episode of star formation is
\begin{equation}
  \label{eq:sn_blowout}
  M_{\rm blow-out} =
 \left[4 \,N_{100}\,\epsilon_{\rm SN}  \left (\frac{V_{\rm out}}{500 \,\kms}
    \right)^{-2}\right]\,\mstar\,.
\end{equation}
The quantity in brackets is the mass-loading factor $\eta$, and is expected
to be of order $\eta=1-2$ for energy-driven winds and $\eta \propto \vmax^{-1}$
(though not exceeding $\sim 5$) for momentum-driven winds (e.g.,
\citealt{martin1999, springel2003, murray2005,
  dave2011}).  

A outflow occurring on a time scale that is short relative to a halo's dynamical
time effectively imparts an impulsive change to energy of the dark matter
particles on scales small relative to the characteristic size of the pre-outflow
gas distribution.  We model this effect by numerically computing an isotropic,
spherically symmetric equilibrium distribution function $f(E)$ of a dark matter
halo, modeled as a \citet{hernquist1990} profile, in the presence of a central
gaseous component that is modeled as a generalized Hernquist profile
(\citealt{dehnen1993, tremaine1994}; see \citealt{boylan-kolchin2007} for a
demonstration of the numerical stability of these models).  The effect of
impulsive blow-out of gas is modeled by removing the gaseous component and
allowing the dark matter component to dynamically evolve to a new equilibrium in
an $N$-body simulation.  This may be viewed as an instantaneous blow-out
scenario, similar to that envisioned by \citet{navarro1996a} and
\citet{read2005}.  As our experiments do not include any contraction of the dark
matter halo in response to the initial assembly of the gaseous component, they
should provide upper limits on the re-arrangement of dark matter in the blow-out
model.  We adopt a host halo of $\vmax=35\,\kms$ with $\rmax=2\,\kpc$ and
investigate removing either $2.2 \times 10^{7}$ or $1.1 \times 10^{8}\,\msun$ of
gas.  The scale radius of the gaseous component is set to 200 pc, and the dark
matter halos are resolved with $2 \times 10^{6}$ particles.

%%%%%%%%%%%%%%%%%%%%%%%%%%%%%%%%%%%%%%%%%%%%%%%%%%%%%%%%%%%%%%
\begin{figure}
 \centering
 \includegraphics[scale=0.55]{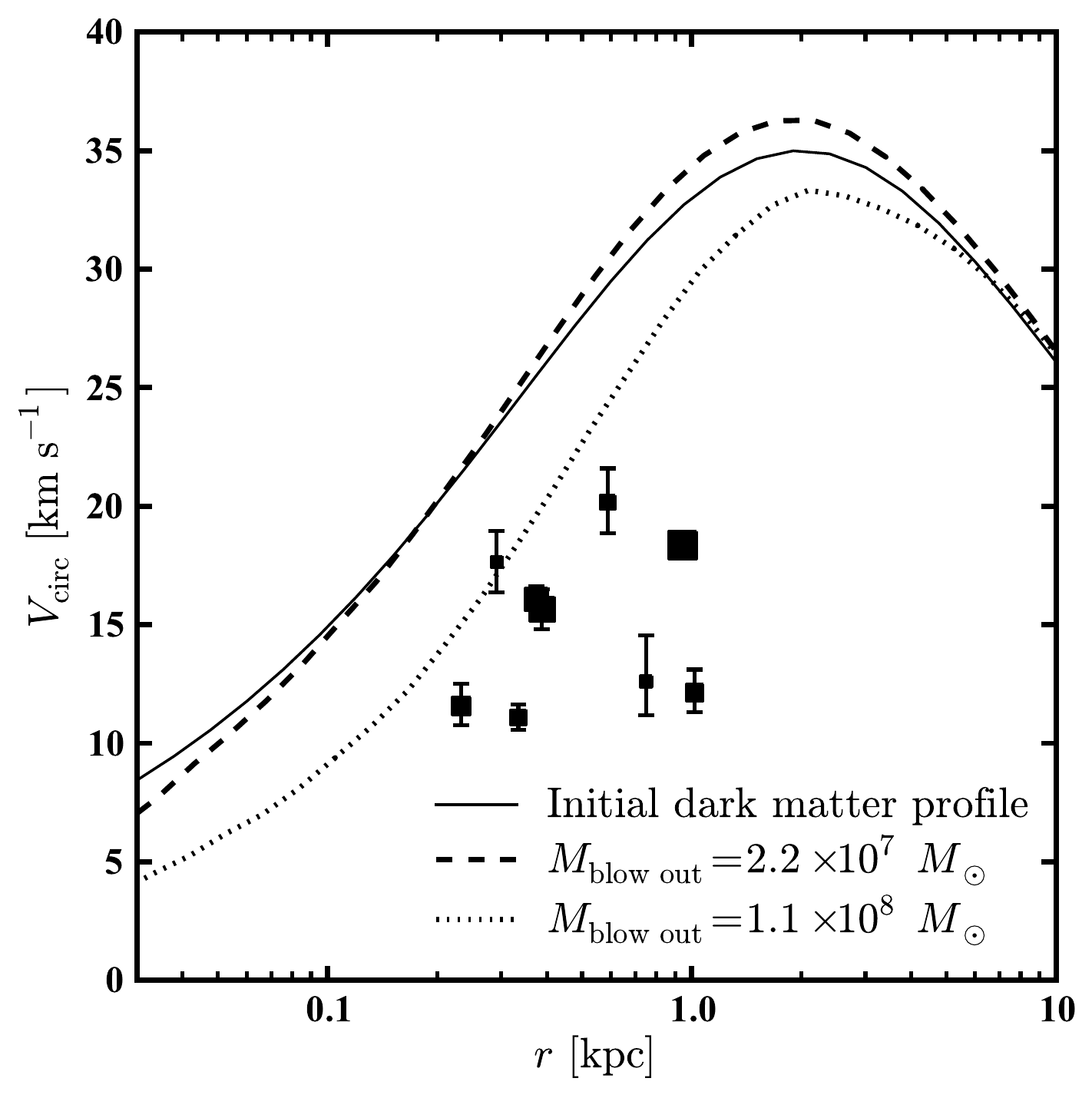}
 \caption{Simulations of impulsive blow-out from a $\vmax=35\,\kms$ halo.  The
   initial halo profile is plotted as a solid line; the dashed and dotted curves
   correspond to the final, relaxed profile after gas blow-out of $2.2 \times
   10^{7}$ and $1.1 \times 10^{8}\,\msun$, respectively.  Measured values of
   $\vcirc$ for the bright dSphs are plotted as squares, sized proportional to
   $\log\,L_V$, with error bars.  We emphasize that the two data points closest
   to the halo line post-blowout are among the least luminous dwarfs we consider
   (Draco and Ursa Minor, with $L_V \simeq 2-4 \times 10^5$).  Matching their
   densities via impulsive feedback would then require ejecting $\sim 100$ 
   times as much mass as is present in stars today in these systems.
\label{fig:blowout_vcirc}
}
\end{figure}
%%%%%%%%%%%%%%%%%%%%%%%%%%%%%%%%%%%%%%%%%%%%%%%%%%%%%%%%%%%%%% 

Figure~\ref{fig:blowout_vcirc} shows the results of these experiments.  Blow-out
causes an immediate drop in the gravitational potential at the system's center.
Particles with $r \la R_{\rm s, gas}$ lose a non-negligible amount of binding
energy and move to larger apocenter distances on a dynamical time-scale, while
particles with pericenters $r_p \gg R_{\rm s, gas}$ are unaffected.  The net
effect is a reduction of the dark matter density for $r \la R_{\rm s, gas}$.
Impulsively removing a gas mass of $\sim 2 \times 10^7\,\msun$ from the center
of a typical $\vacc=35\,\kms$ halo has a negligible effect on the dark matter
density profile (dashed line), while removing $\sim 10^8\,\msun$ of gas reduces
the central $\vcirc$ by approximately 25\%.  (The inner density cusp slope,
which is originally $-1$, ends up closer to $-0.5$ after this maximal blow-out.)
This result is consistent with \citet{navarro1996a}, who used numerical
simulations to show that a supernova-driven outflow following a single episode
of star formation that created $\sim 10^{8}\,\msun$ of stars can create a core
in the dark matter halo of a $\vcirc \sim 50\,\kms$ dwarf galaxy.  It also
agrees with the work of \citet{read2005}, which found that removing the vast
majority of a dwarf galaxy's baryons (95-99\%) in either one or two impulsive
episodes significantly flattens a dark matter density cusp, with two blow-outs
creating a dark matter core.  \citet{gnedin2002} demonstrated that removal of a
gaseous disk containing nearly all of a dwarf galaxy's baryons will reduce the
central dark matter density by approximately 50\%; our reduction of $\sim 25\%$
in $\vcirc$ is in good agreement with this result.

The final dark matter profile after removing $10^{8}\,\msun$ of gas agrees only
with Draco among the dSphs (square symbols with error bars).  Blowing out
$10^8\,\msun$ of gas requires forming $\eta^{-1}\,10^{8}\,\msun$ of stars,
however [see Eqn.~\eqref{eq:sn_blowout}].  Since the bright dSphs have $\mstar
\approx 5\times 10^{5}-5\times 10^{7}\,\msun$ and $\eta$ should be of order
unity (and certainly not in excess of 10), only the brightest dSph (Fornax) has
formed enough stars to remove the necessary amount of gas.  In particular, the
necessary amount of star formation exceeds the stellar content of Draco by a
factor of $\sim 200$, and the reduction in density is not sufficient to match
Fornax's observed $\vcirc$ (the largest square, with $\vcirc \sim 17\,\kms$ at
$\sim 0.9\,\kpc$).  

Starting with a much more concentrated baryonic component -- e.g., with a scale
radius of 1-10 pc rather than 100 pc, as assumed in this work -- could cause gas
blow-out to have a more deleterious effect on the central region of the dark
matter halo.  Collecting $10^8\,\msun$ of gas in such a small region would
likely steepen the initial dark matter profile a great deal, however.  It would
also result in star formation that is very centrally concentrated, in contrast
to the typical half-light radii of the bright dwarfs of $300 \la \rhalf \la
1000\,{\rm pc}$.  If the stars are indeed formed at very small radii, they would
have to migrate outward in order to match observed sizes of the dSphs, which
would perhaps result in radially biased velocities in the dwarfs' central
regions.  We conclude that absent outflows that are much larger than expected in
usual models of galaxy formation (i.e., absent mass loading factors of
$\eta=10-100$) or an extremely concentrated gaseous distribution, impulsive gas
removal due to supernova feedback is likely insufficient to lower the central
densities of the massive subhalos predicted by \lcdm\ enough to agree with
observations of the bright dSphs.  In the future, it may be possible to more
directly constrain the amount of mass removed from dSphs through detailed
analysis of chemical abundance patterns \citep{kirby2011b}.

\citet{governato2010} have recently performed a very high resolution simulation
with a star formation density threshold of $100 \,{\rm cm}^{-3}$ and found it
produces a cored galaxy with $\mstar = 5 \times 10^{8}\,\msun$ in a halo of
$\mvir=5.7 \times 10^{10}\,\msun$ ($\vcirc \sim 60 \,\kms$).
\citet*{pontzen2011} argued that the effects of repeated cycles of star
formation, blow-out, and re-cooling can explain these results.  Unless this
process is somehow much more efficient than the instantaneous blow-out scenario
discussed above -- i.e., unless it removes much more gas mass per stellar mass
formed than does a single episode of instantaneous blow-out -- it is unlikely to
explain the low densities of the Milky Way dwarf spheroidals: the Governato et
al. galaxy has a stellar mass that is a factor of 10 larger than any of the MW
dSphs (and a factor of 1000 larger than several of the bright dSphs).  High
resolution hydrodynamical simulations of the Milky Way's satellites find that
the net effect of star formation and feedback is either negligible or causes an
increase in the dark matter density \citep{di-cintio2011, parry2012}. Clearly,
it will be invaluable to have a larger number of simulations of possible dSphs
over a wider mass range, simulated with different codes and stellar feedback
prescriptions.  It will also be interesting to better constrain the possible
role of black hole feedback, which can power strong outflows without relying on
star formation, in the Milky Way dSphs.  \citet{reines2011} have recently
discovered an AGN in a dwarf starbursting galaxy with a mass similar to that of
the LMC (i.e., more massive than the MW dSphs); as of yet, there is no evidence
for central black holes in the dSphs, with \citet{jardel2012} finding a
$1\,\sigma$ upper limit of $3.2 \times 10^4\,\msun$ for the mass of a potential
black hole at the center of Fornax.

\subsection{Stochastic galaxy formation?}
\label{subsec:stochastic}
If baryonic feedback has not strongly modified the structure of massive MW
subhalos, and the abundance of these objects is commensurate with that found in
the Aquarius simulations, then it seems unavoidable that galaxy formation must
be \textit{highly} stochastic in halos of $\la 50\,\kms$ ($\macc \la
10^{10}\,\msun$).  By this, we mean that the stellar mass of a galaxy cannot
correlate with its dark matter halo mass at these scales (see
Fig.~\ref{fig:eff_macc}).  Stochasticity is not unexpected in low halo masses,
as gas cooling will depend strongly on the ability to form molecular gas, which
in turn depends on the gas metallicity \citep{kuhlen2011}.  An additional source
of scatter in stellar mass at fixed halo mass comes from allowing satellites to
have a $\mhalo-\mstar$ relation that differs from that of central galaxies
\citep{neistein2011a}. \citet{yang2011a} have demonstrated that such a variation
is expected because the stellar mass of a satellite should be related to the
$\mhalo-\mstar$ relation at $\zacc$, not at $z=0$.  (It remains to be seen
whether the abundance of MW satellites is correctly reproduced in this or
similar abundance matching models.)

However, the stochasticity required by the results of
Table~\ref{table:posteriors} and Fig.~\ref{fig:eff_macc} is somewhat more
curious: it requires a strong (and systematic) suppression of galaxy formation
in the most dense subhalos of the Milky Way, which is counter-intuitive.  It
also requires that the luminosities of \textit{all} subhalos with $30 \la \vmax
\la 60\,\kms$ -- a range in which we expect to find several objects in the Milky
Way -- host galaxies with luminosities a factor of 10-1000 lower than subhalos
in the range $15 \la \vmax \la 30\,\kms$.  If this is due to scatter, it must be
only scatter in one direction, toward low $\mstar$ at fixed $\mhalo$, in the
mass range of massive failures, which would host galaxies with luminosities
comparable to ultra-faint dwarf spheroidal galaxies in this scenario.

The dynamics of the ultra-faint are not as well constrained as those of the
bright dSphs for two main reasons.  First, the kinematic data for the
ultra-faint galaxies of the Milky Way are not of the same quality as for the
bright dSphs.  Additionally, the ultra-faints have smaller half-light radii,
which means that measurements of $\mhalf$ have less power in constraining
$\vmax$.  It is therefore possible that some, or all, of the massive failures
are accounted for by ultra-faint galaxies; at least, this is not presently ruled
out by the data.  If this is the case, however, then galaxy formation must be
\textit{extremely} stochastic in halos of $\vacc \la 40-50\,\kms$, as it would
require $M/L$ variations in excess of 1000 at this scale to simultaneously
explain the ultra-faints and the brightest dSphs.  As shown in
Fig.~\ref{fig:mfilter_30}, the massive failures are all more massive than the
photo-suppression threshold at all times, so it is not clear what shuts off star
formation in the massive failures hosting ultra-faints in this scenario.  For
example, the hybrid simulations of \citet{bovill2011, bovill2011a}, which
reproduce many observed properties of the ultra-faint dwarf spheroidals by
placing them in halos below the filtering mass, over-produce brighter Milky Way
dwarfs, even though only star formation prior to reionization is considered.

\subsection{Different dark matter physics on sub-galactic scales?}
\label{subsec:dm_physics}
Should none of the explanations in Sections~\ref{subsec:mw_mass} -
\ref{subsec:stochastic} prove correct, then modifications of the \lcdm\ model on
small scales may be necessary to explain the structure of the MW dSphs.  Indeed,
\citet{lovell2012} have already examined the structure of a version of the
Aquarius A halo simulated in a Warm Dark Matter cosmology.  They find that WDM
substantially alleviates the massive failure problem, as the subhalos in this
simulation are (1) less numerous and (2) less dense than their counterparts in
the CDM run.  Although this result is very promising, it remains to be seen
whether WDM simulations produce \textit{enough} substructure to account for the
ultra-faint galaxies of the MW \citep{maccio2010a, polisensky2011}.  The
particle mass used by \citet{lovell2012} is also near the lower limit of what is
allowed by observations of Lyman-$\alpha$ absorption features in the spectra of
distant quasars \citep{boyarsky2009}.  Improved measurements of this type will
further constrain the allowed parameter space for WDM and the viability of
$\Lambda$WDM as a cosmological model.

Alternatively, it may be the case that dark matter is not strictly
collisionless, but rather has a non-negligible cross section for self-scattering
\citep{spergel2000}.  This self-interacting dark matter (SIDM) model was
originally proposed to resolve the missing satellites problem and to explain the
apparently overdense hosts of low surface brightness galaxies predicted by CDM,
but interest in SIDM mostly abated when it was shown that SIDM would produce
strong effects in clusters of galaxies (among other reasons; see, e.g.,
\citealt{yoshida2000a}).  Nevertheless, the range of models explored was
relatively restricted, and \citet*{loeb2011} have recently argued that SIDM with
a velocity-dependent cross section peaking at a velocity scale relevant for
dwarfs retains the benefits of original SIDM models while avoids their drawbacks
(see also \citealt{feng2009}).  Simulating structure formation in such
a model would therefore be of significant interest.

\subsection{Missing Physics at $\bvfifty$?}
\label{subsec:missing_physics}
%%%%%%%%%%%%%%%%%%%%%%%%%%%%%%%%%%%%%%%%%%%%%%%%%%%%%%%%%%%%%%
\begin{figure}
 \centering
 \includegraphics[scale=0.55]{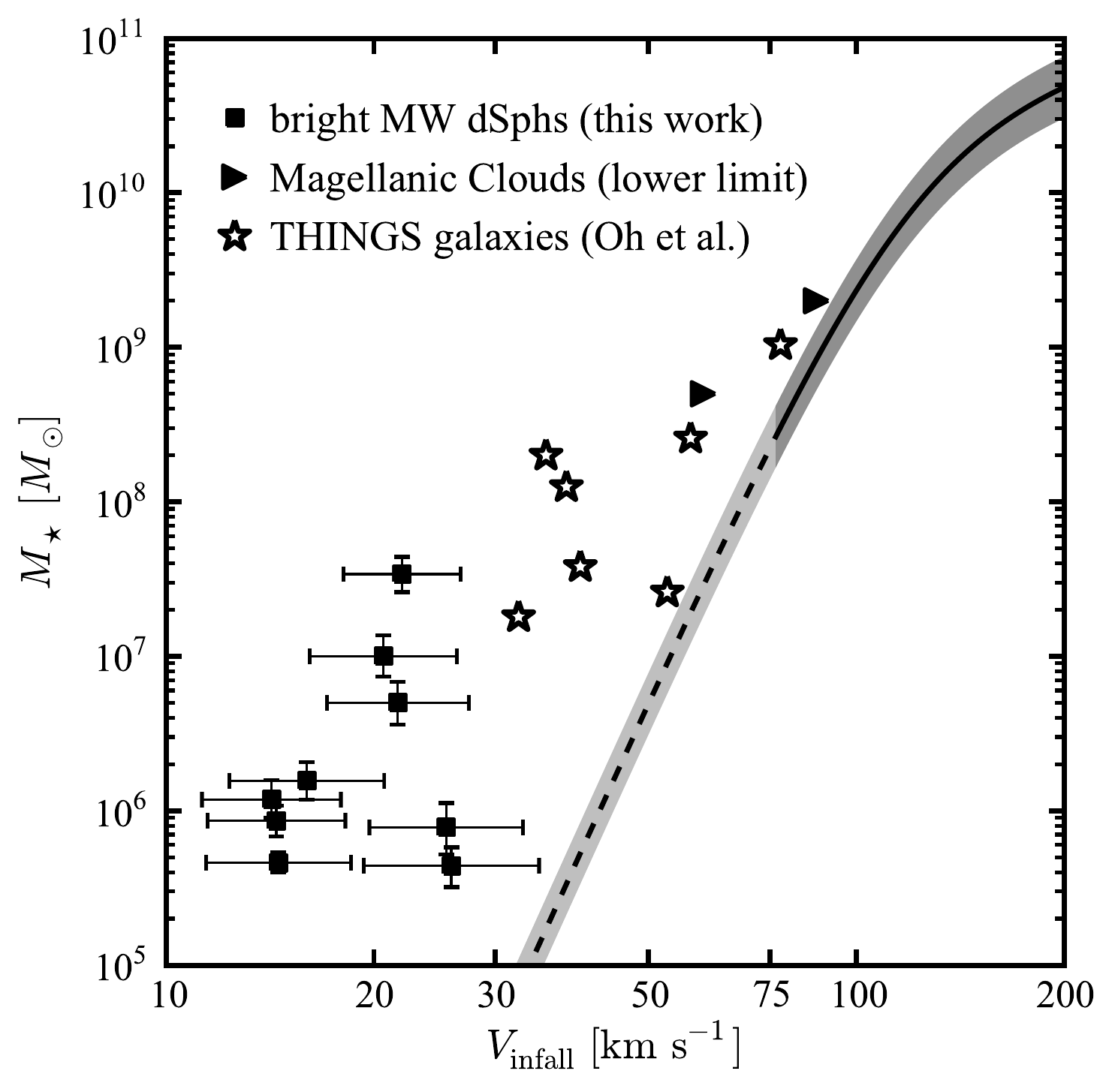}
 \caption{Inferred $\mstar(\vacc)$ relation for bright MW dSphs (black squares
   with error bars, based on calculations in Sec.~\ref{subsec:posteriors}).  The
   Magellanic Clouds (right-pointing triangles) are placed on the plot at their
   current values of $V_{\rm flat}$, which is a lower limit to $\vacc$.
   Observations of low-mass field galaxies from THINGS (tabulated in
   \citealt{oh2011a}) are plotted as open black stars.  These galaxies all lie
   higher than the $z=0$ abundance matching relation (solid curve), as well as
   its extrapolation to lower 
   $\vacc$ (dashed curve), and the deviations are systematically larger at lower
   values of $\vacc$.  The shaded region around the abundance matching relation
   shows a scatter of 0.2 dex in $\mstar$ at fixed $\vacc$, which is the upper
   limit allowed for massive halos $\vacc \ga 150\,\kms$ \citep{guo2010,
     behroozi2010}.
 \label{fig:vacc_mstar_things}
}
\end{figure}
%%%%%%%%%%%%%%%%%%%%%%%%%%%%%%%%%%%%%%%%%%%%%%%%%%%%%%%%%%%%%% 

Ultimately, the structure of the Milky Way dwarfs must be understood as part of
a full theory of galaxy formation that can explain the formation, evolution, and
properties of all galaxies.  It is therefore reasonable to ask whether the
structural issues we have raised regarding MW dSphs are also seen in other
systems of similar mass.  Although observing galaxies at these masses over
cosmological volumes is essentially impossible at present, THINGS 
\citep{walter2008} has observed 34 nearby galaxies ($D < 15\,\mpc$) in HI with
high spatial resolution.  We use the results of Oh et al. (2011a,b\nocite{oh2011,
  oh2011a}), who performed a detailed mass modeling of seven low-mass THINGS
galaxies, to compare the $\mstar - \vacc$ relation for these isolated galaxies
to the bright MW dSphs and to abundance matching expectations in
Fig.~\ref{fig:vacc_mstar_things}.  

Intriguingly, the THINGS data mostly lie above the abundance matching relation
and its extrapolation to lower masses and provide a smooth transition from the
abundance matching $\mstar - \vacc$ relation at $\vacc \ga 75 \,\kms$ to the
regime of the MW dSphs at $\sim 15-25\,\kms$ derived in this work.  Simulations
of isolated dwarf galaxies with masses comparable to the THINGS sample also fall
above the abundance matching curve \citep{sawala2011}.  Note, however, that if
the \lcdm\ model is correct, and galaxies with $\vacc \approx 40\,\kms$
typically have stellar masses of $\sim 10^{8}\,\msun$, then current stellar mass
functions are substantially underestimating the abundance of such galaxies.
Alternatively, either most halos with $\vacc \approx 40\,\kms$ host systems with
much lower total stellar masses, or the abundance of halos at these masses is
lower than what is predicted by \lcdm.  There are at least three additional
lines of evidence arguing that isolated halos with $\vcirc \approx 50\,\kms$ do
not match \lcdm\ expectations:
\begin{itemize}
\item \textbf{HI observations}: The ALFALFA survey has performed a blind 21-cm
  emission line search over a wide area to look for neutral hydrogen in
  galaxies.  \citet{papastergis2011} have shown that while the velocity width
  function $\Phi_w$ measured from ALFALFA agrees fairly well with \lcdm\
  predictions for massive galaxies, the observed number counts fall below those
  predicted by \lcdm\ for $w \la 100\,\kms$ (corresponding approximately to
  $\vmax \la 75 \,\kms$, assuming an average conversion of $w_{50}=0.75 \,\vmax$).
  The discrepancy reaches a factor of $\sim\!8$ at $w=50\,\kms$ ($\vmax \approx
  37\,\kms$) and becomes even worse at lower $w$.

\item \textbf{Void galaxies}: \citet*{tikhonov2009} analyzed properties of voids
  in the Local Volume in comparison to theoretical predictions.  They find that
  the abundance of void galaxies is over-predicted by a factor of $\sim\!10$ in
  \lcdm\ at $\vcirc \approx 40\,\kms$, and that the void size distribution is
  only reproduced if halos of $\vcirc \la 40\,\kms$ do not host void galaxies
  (but see \citealt{tinker2009}).

\item \textbf{Damped Lyman-$\bm{\alpha}$ systems}: The gaseous content of dark
  matter halos at $z \sim 3$ can be probed by quasar absorption
  spectra. \citet{barnes2009} have shown that many of the properties of damped
  Lyman-$\alpha$ systems can be understood in \lcdm-based models.  This success
  comes at the expense of requiring halos with $\vcirc \la 50 \,\kms$ to be very
  baryon-poor.  As noted in Section~\ref{subsec:high_z}, $50\,\kms$ is well
  above the photo-suppression scale at this redshift, indicating that
  reionization should not have caused such halos to lose a substantial amount of
  their baryons.
\end{itemize}

There are, of course, many potential sources of these disagreements, and the
underlying \lcdm\ theory is certainly not the most likely of of these sources.
A better understanding of feedback from star formation and its effects on halos
of $\vacc \la 50\,\kms$ will be crucial, and may explain all of these apparent
discrepancies, as well as other issues such as the central densities of low
surface brightness galaxies (though see \citealt{kuzio-de-naray2011} for
arguments against baryonic physics explaining the density structure of these
galaxies).  It is imperative not to rely on plausibility arguments for the
effects of feedback, but rather to understand whether realistic feedback models
can actually produce dwarf spheroidal galaxies with properties akin to those
seen in the Milky Way (as challenging as this may be!).

\section{Conclusions}
\label{sec:conclusions}
In this paper, we have expanded on the arguments of \bkbk, where we first showed
that the bright satellites of the Milky Way apparently inhabit dark matter
subhalos that are substantially less dense that the most massive subhalos from
state-of-the-art \lcdm\ simulations.  Using subhalo profiles computed directly
from the simulations rather than assuming subhalos are fit by NFW profiles, we
have confirmed our previous result.  Furthermore, we have now computed the most
likely $\vmax$, $\vacc$, and $\macc$ values of the dwarf spheroidals using a
likelihood analysis of the Aquarius data.  This procedure predicts that all of
the Milky Way dwarf spheroidals reside in halos with $\vmax \la 25\,\kms$,
whereas more than ten subhalos per host halo are expected to have $\vmax > 25
\,\kms$.

This ``massive failure'' problem cannot be solved by placing the bright
satellites in the subhalos with the largest values of $\vmax$ at infall or at
the epoch of reionization: as discussed in Section~\ref{subsec:high_z} and shown
in Figures~\ref{fig:mfilter_30} -- \ref{fig:massfunc_z6}, the missing subhalos
are among the most massive at all previous epochs as well.  Explaining this lack
of galaxies in the expected massive subhalos is not natural in standard
\lcdm-based galaxy formation models: options include (1) a Milky Way halo that
either is significantly deficient in massive subhalos, or is populated by
subhalos with much lower concentrations than are typical; (2) stochasticity in
galaxy formation at low masses, such that halo mass and luminosity have
essentially no correlation; (3) strong baryonic feedback that reduces the
central density of all massive subhalo by a large amount ($\ga 50\%$ reduction
on scales of $\sim 0.5$ kpc).

We have argued above that these solutions all seem fairly unlikely as individual
causes.  It might be possible to apply them all at once: if the Milky Way halo
mass sits at the low end of current constraints ($\sim 10^{12}\,\msun$),
\textit{and} galaxy formation produces order unity scatter in $\mstar$ at fixed
halo mass below $\sim 50 \,\kms$, \textit{and} baryonic feedback is able to
alter the central densities of dark matter halos in a maximal way, then it may
be possible to explain the low densities of the MW dSph galaxies.  We find this
combination somewhat implausible, but it is certainly worth exploring.  A
detailed comparison of the masses of M31 dSph galaxies will be particularly
useful in assessing the "rare Milky Way" hypothesis.  Finally, if we do reject
solutions (1)-(3), then we are left with the question of the nature of dark
matter.  Allowing for phenomenology such as self-interactions, decays, or
non-negligible thermal velocities may explain the puzzles discussed here without
destroying the many successes of cold dark matter models on large scales.

\vspace{0.1cm}

\section*{Acknowledgments} 
We thank Carlos Frenk, Fabio Governato, Martin Haehnelt, Amina Helmi, Juna
Kollmeier, Andrey Kravtsov, Erik Tollerud, and Simon White for helpful and spirited
discussions.  The Aquarius Project is part of the programme of the Virgo
Consortium for cosmological simulations; we thank the Aquarius collaboration for
giving us access to their simulation data. MB-K thanks Takashi Okamoto for
providing the data for $M_c(z)$ plotted in Fig.~\ref{fig:mfilter_30}.  The
Millennium and Millennium-II Simulation databases used in this paper were
constructed as part of the activities of the German Astrophysical Virtual
Observatory.  MB-K acknowledges support from the Southern California Center for
Galaxy Evolution, a multi-campus research program funded by the University of
California Office of Research.  JSB was supported by NSF AST-1009973; MK was
supported by NASA grant NNX09AD09G.

\bibliography{draft}
\appendix
\section{Mass profiles of simulated subhalos}
\label{sec:appendix}
In \bkbk, we used the values of $\vmax$ and $\rmax$ computed in
\citet{springel2008} to construct NFW profiles for the Aquarius subhalos.  While
subhalos are generally well-fitted by NFW profiles in the aggregate, any
individual subhalo may show (significant) deviations from the NFW profile
specified by its $\{\vmax, \,\rmax\}$ values.  In this paper, therefore, we use
the raw particle data to compute the circular velocity profiles of the Aquarius
subhalos.

While using the raw particle data provides a much better estimate of the true
mass profile of a subhalo than does the NFW assumption, we still must remember
that we are interested in properties of subhalos on scales that are not much
larger than the force resolution of the Aquarius simulations.  Draco, for
example, has a de-projected half-light radius of 291 pc \citep{wolf2010}, while
the Plummer-equivalent softening length $\epsilon$ of the Aquarius simulations
is 65 pc.  The effect of force softening is to reduce the density on scales of
$\sim 3\,\epsilon$ or smaller (note that forces are Newtonian on scales larger
than $2.8\,\epsilon$).  This will not affect differential quantities such as
$\rho(r)$ on larger scales, but will affect cumulative properties such as
$\vcirc(r)$, such that the measured mass is an underestimate of the true mass.
For example, \citet{font2011} find that the mass within 300 pc of the centers of
Aquarius A-2 subhalos (the resolution level of the simulations used in this
paper) is systematically underestimated by approximately 20\% by comparing with
the higher resolution Aquarius A-1 simulation.

\begin{figure}
 \centering
\includegraphics[scale=0.58]{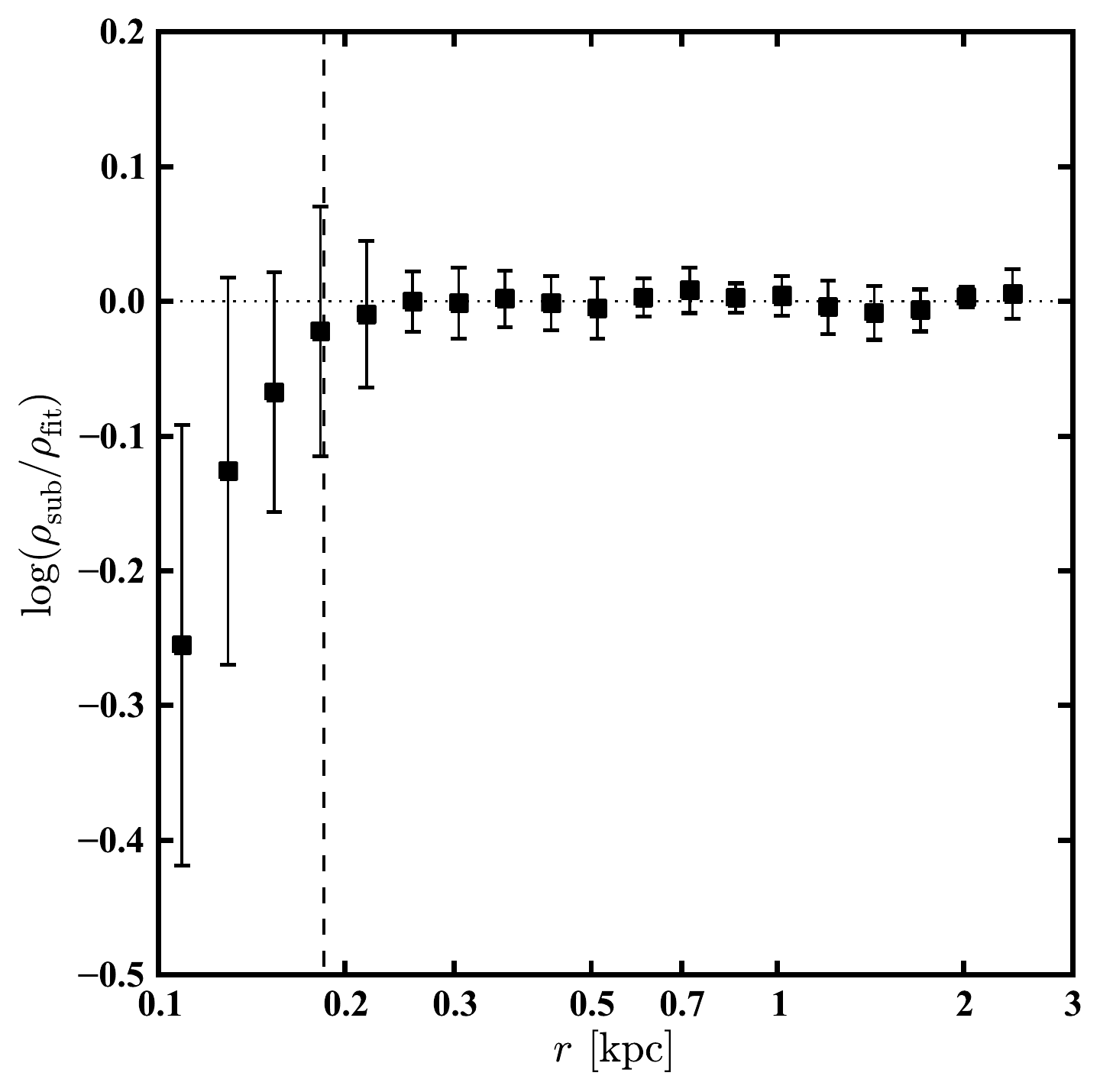}
\caption{The difference between the measured subhalo density profiles and the
  best-fitting Einasto density profiles as a function of radius.  The mean
  values (symbols), along with the standard deviations (error bars) based on the
  ten highest $\vmax$ subhalos in the Aquarius B halo are plotted.  The subhalos
  follow Einasto profiles (all with goodness-of-fit $Q$ values 
  of $<0.03$), with very small residual scatter, for $r \ga 250\, {\rm pc}$.  On
  smaller scales, however, the measured profile falls below the Einasto fit
  because of gravitational softening.  This causes an underestimate of
  $\vcirc(r)$ for  $r \la 2.8\,\epsilon$ 
  (forces for $r > 2.8 \,\epsilon$ are exactly Newtonian); this
  value is marked with a vertical dashed line.
 \label{fig:ein_fit}
}
\end{figure}

We adopt the following procedure to correct for the effects of force softening.
First, we fit the density profile of each subhalo with an Einasto profile on the
range\footnote{We choose 291 pc for the lower limit of our fits because this is
  the de-projected half-light radius of Draco, the most dense dSph in our bright sample.}
$[291 \, {\rm pc}, r_{\rm upper}]$, where $r_{\rm upper}$ is the smaller of 3
kpc or $1.5\,\rmax$.  This procedure typically results in very good fits: using
the goodness-of-fit measure
\begin{equation}
  \label{eq:ein_fit}
  Q^2=\frac{1}{N_{\rm bins}} \,\sum_{i=1}^{N_{\rm bins}}[\log \rho_{\rm
    subs}(r_i) - \log \rho_{\rm model}(r_i)]^2 \,,
\end{equation}
we find a median (mean) $Q$ value of 0.057 (see also Fig.~\ref{fig:ein_fit}).
We then use the best-fitting Einasto profile to model the density distribution
for $r < 291$ pc and the raw particle distribution for $r \ge 291$ pc.  The
correction to $\vcirc$ from this procedure will be of decreasing importance with
increasing $r$; the correction is typically 5-10\% at 291 pc.  For a small
number of subhalos (typically with low numbers of particles), the best-fitting
Einasto profile has $Q > 0.1$.  In this case, we opt to be conservative and do
not apply any correction to the measured particle distribution.  We also do not
apply the correction for subhalos with $\rmax < 500\,{\rm pc}$.
%%%%%%%%%%%%%%%%%%%%%%%%%%%%%%%%%%%%%%%%%%%%%%%%%%%%%%%%%%%%%%
%%%%%%%%%%%%%%%%%%%%%%%%%%%%%%%%%%%%%%%%%%%%%%%%%%%%%%%%%%%%%

To assess the impact of this softening correction on our results, we re-ran the
analysis from  Section~\ref{subsec:posteriors} without including the correction;
Table~\ref{table:posteriors_append} compares the two sets of results.  The
softening correction has minimal impact on the derived values of $\vmax$, with
none of the dSphs changing by more than 5\%.  $\vacc$ and $\macc$ are very
stable as well: only Draco, the dSph with both the smallest half-light radius
and largest inferred $\vmax$, is affected at the 9\% level in $\vacc$ (20\% in
$\macc$).  Even these changes are much smaller than the errors on these values.
We therefore conclude that our results are not driven by including (or
excluding) the correction for gravitational softening.
%%%%%%%%%%%%%%%%%%%%%%%%%%%%%%%%%%%%%%%%%%%%%%%%%%%%%%%%%%%%%
\begin{table}
  \caption{
    Effects of the softening correction on derived properties of the Milky Way dwarfs.  For each
    dSph, we list two sets of numbers: the first row includes the softening
    correction, while the second row does not include the softening correction.
    \label{table:posteriors_append}
  }
  \begin{tabular}{lcccc}
    \hline
    \hline
    Name & & $\vmax$ & $\vacc$ & $\macc$\\
    \Bspace & & [$\kms$] & [$\kms$] & [$\msun$]\\
    \hline
Fornax \Tspace \Bspace & & $17.8^{+0.7}_{-0.7}$ &  $22.0^{+4.7}_{-3.9}$ &
$7.4^{+6.1}_{-3.3} \times 10^{8}$\\   
 \Tspace \Bspace & & $17.9^{+0.6}_{-0.6}$ &  $21.8^{+4.2}_{-3.5}$ & $7.2^{+5.1}_{-3.0} \times 10^{8}$ \\ 
Leo I  \Tspace \Bspace & & $16.4^{+2.3}_{-2.0}$ &  $20.6^{+5.7}_{-4.5}$ & $5.6^{+6.8}_{-3.1} \times 10^{8}$ \\    
 \Tspace \Bspace & & $16.6^{+1.8}_{-1.6}$ &  $20.9^{+4.8}_{-3.9}$ & $5.7^{+5.5}_{-2.8} \times 10^{8}$ \\ 
Sculptor  \Tspace \Bspace & & $17.3^{+2.2}_{-2.0}$ &  $21.7^{+5.8}_{-4.6}$ & $6.6^{+7.8}_{-3.6} \times 10^{8}$ \\
 \Tspace \Bspace & & $17.6^{+1.7}_{-1.5}$ &  $22.1^{+4.5}_{-3.7}$ & $6.9^{+6.0}_{-3.2} \times 10^{8}$ \\ 
Leo II  \Tspace \Bspace & & $12.8^{+2.2}_{-1.9}$ &  $16.0^{+4.7}_{-3.6}$ & $2.4^{+3.1}_{-1.4} \times 10^{8}$ \\
 \Tspace \Bspace & & $13.2^{+2.1}_{-1.8}$ &  $16.7^{+4.3}_{-3.4}$ & $2.7^{+3.3}_{-1.5} \times 10^{8}$ \\ 
Sextans  \Tspace \Bspace & & $11.8^{+1.0}_{-0.9}$ &  $14.2^{+3.7}_{-2.9}$ & $1.9^{+1.7}_{-0.9} \times 10^{8}$\\  
 \Tspace \Bspace & & $11.9^{+0.9}_{-0.9}$ &  $14.3^{+3.7}_{-3.0}$ & $1.9^{+1.7}_{-0.9} \times 10^{8}$ \\
Carina  \Tspace \Bspace & & $11.4^{+1.1}_{-1.0}$ &  $14.4^{+3.7}_{-3.0}$ & $1.8^{+1.8}_{-0.9} \times 10^{8}$ \\  
 \Tspace \Bspace & & $11.4^{+1.0}_{-0.9}$ &  $14.5^{+3.5}_{-2.8}$ & $1.8^{+1.7}_{-0.9} \times 10^{8}$ \\ 
Ursa Minor  \Tspace \Bspace & & $20.0^{+2.4}_{-2.2}$ &  $25.5^{+7.4}_{-5.8}$ & $1.1^{+1.5}_{-0.6} \times 10^{9}$ \\
 \Tspace \Bspace & & $20.3^{+2.2}_{-2.0}$ &  $26.1^{+7.4}_{-5.8}$ & $1.2^{+1.6}_{-0.7} \times 10^{9}$ \\ 
Canes Venatici I  \Tspace \Bspace & & $11.8^{+1.3}_{-1.2}$ &  $14.5^{+4.0}_{-3.1}$ &
$1.9^{+2.0}_{-1.0} \times 10^{8}$\\    
 \Tspace \Bspace & & $11.8^{+1.3}_{-1.2}$ &  $14.6^{+4.0}_{-3.1}$ & $1.9^{+2.0}_{-1.0} \times 10^{8}$ \\  
Draco  \Tspace \Bspace & & $20.5^{+4.8}_{-3.9}$ &  $25.9^{+8.8}_{-6.6}$ &
$1.2^{+2.0}_{-0.7} \times 10^{9}$\\ 
 \Tspace \Bspace & & $21.6^{+4.0}_{-3.4}$ &  $28.2^{+8.3}_{-6.4}$ & $1.5^{+2.1}_{-0.9} \times 10^{9}$ \\  
\hline
\end{tabular}
\end{table}
%%%%%%%%%%%%%%%%%%%%%%%%%%%%%%%%%%%%%%%%%%%%%%%%%%%%%%%%%%%%%

\label{lastpage}
\end{document}